\documentclass[%
 reprint,
superscriptaddress,
 amsmath,amssymb,
 aps,
pra,
]{revtex4-2}

\usepackage{graphicx}
\graphicspath{{./figs/}}
\usepackage{caption}

\usepackage{subcaption}
\usepackage{dcolumn}
\usepackage{bm}
\usepackage{algorithm,algcompatible}
\usepackage[usenames,dvipsnames]{color} 
\usepackage{ifthen} 
\usepackage{braket}
\usepackage{amsmath}
\usepackage{amssymb}
\usepackage[toc,page]{appendix}

\def\shownoteal{1} 
\newcommand{\nal}[1]{\ifthenelse{\shownoteal=1}{\textcolor{red}{[[#1]]}}{}}

\begin{document}

\preprint{APS/123-QED}

\title{Neural Networks for Programming Quantum Annealers}

 \author{Samuel Bosch}
 \email{sbosch@mit.edu}
 \address{
 Massachusetts Institute of Technology, Cambridge, MA 02139, USA\\
}

\author{Bobak Kiani}
\address{
 Massachusetts Institute of Technology, Cambridge, MA 02139, USA\\
}
\author{Rui Yang}
\address{
University of Waterloo, Waterloo,
Ontario, N2L 3G1, Canada\\
}

\author{Adrian Lupascu}%
\address{
University of Waterloo, Waterloo,
Ontario, N2L 3G1, Canada\\
}

\author{Seth Lloyd}
\address{
 Massachusetts Institute of Technology, Cambridge, MA 02139, USA\\
}

\date{\today}

\begin{abstract}
 Quantum machine learning is an emerging field of research at the intersection of quantum computing and machine learning. It has the potential to enable advances in artificial intelligence, such as solving problems intractable on classical computers. Some of the fundamental ideas behind  quantum machine learning are very similar to kernel methods in classical machine learning. Both process information by mapping it into high-dimensional vector spaces without explicitly calculating their numerical values. Quantum annealers are mostly studied in the adiabatic regime, a computational model in which the quantum system remains in an instantaneous ground energy eigenstate of a time-dependent Hamiltonian. Our research focuses on the diabatic regime where the quantum state does not necessarily remain in the ground state during computation. Concretely, we explore a setup for performing classification on labeled classical datasets, consisting of a classical neural network connected to a quantum annealer. The neural network programs the quantum annealer's controls and thereby maps the annealer's initial states into new states in the Hilbert space. The neural network's parameters are optimized to maximize the distance of states corresponding to inputs from different classes and minimize the distance between quantum states corresponding to the same class. Recent literature showed that at least some of the "learning" is due to the quantum annealer, connecting a small linear network to a quantum annealer and using it to learn small and linearly inseparable datasets. In this study, we consider a similar but not quite the same case, where a classical fully-fledged neural network is connected with a small quantum annealer. In such a setting, the fully-fledged classical neural-network already has built-in nonlinearity and learning power, and can already handle the classification problem alone, we want to see whether an additional quantum layer could boost its performance. We simulate this system to learn several common datasets, including those for image and sound recognition. We conclude that adding a small quantum annealer does not provide a significant benefit over just using a regular (nonlinear) classical neural network. 

\end{abstract}

\maketitle


\section{\label{sec:introduction}Introduction:\protect
}
Machine learning (ML) is among the most exciting prospective applications of quantum technologies. As kernel-based methods have become a pillar of modern ML applications, it is natural to explore their applications within ML applications using quantum computing. The fundamental idea of quantum computing is surprisingly similar to the principle behind kernel methods in ML. That is, to efficiently perform computation in an intractably large vector space. In the case of quantum computers, this vector space is a Hilbert space. Quantum algorithms aim to perform efficient computations in a Hilbert space that grows exponentially with the size of a quantum system. Efficient, in this context, means that the number of operations applied to the system grows at most polynomially with the size of the system.

There are different approaches to building physical quantum computers. The most well-known approaches are universal gate model quantum computers and quantum annealers. Universal gate model quantum computing (or general purpose quantum computing) is universal and flexible. However, building and maintaining the stability of physical qubits is hard. Quantum annealers are a less flexible paradigm of quantum computing but are easier to implement and realize than universal gate model quantum computers. 

Quantum annealing started as a purely theoretical combinatorial optimization method \cite{albash2018adiabatic}. It is mostly used for solving optimization problems formulated in terms of finding ground states of classical Ising spin Hamiltonians \cite{kadowaki1998quantum, crosson2021prospects}. More recently, though, it is also understood as a heuristic optimization method implemented in physical quantum annealing hardware \cite{johnson2011quantum}. A physical implementation of quantum annealing on quantum hardware might lead to speedups over algorithms running on classical hardware in some specific cases. The construction of commercial (non-universal) quantum annealing processors, such as by the Canadian quantum computing company D-Wave Systems \cite{harris2010experimental, berkley2010scalable, berkley2013tunneling, bunyk2014architectural, lanting2014entanglement, dickson2013thermally}, was inspired by earlier theoretical proposals \cite{kaminsky2004scalable, kaminsky2004scalable2}.

\begin{figure*}[htp]
    \centering
    \includegraphics[width=1.0\textwidth]{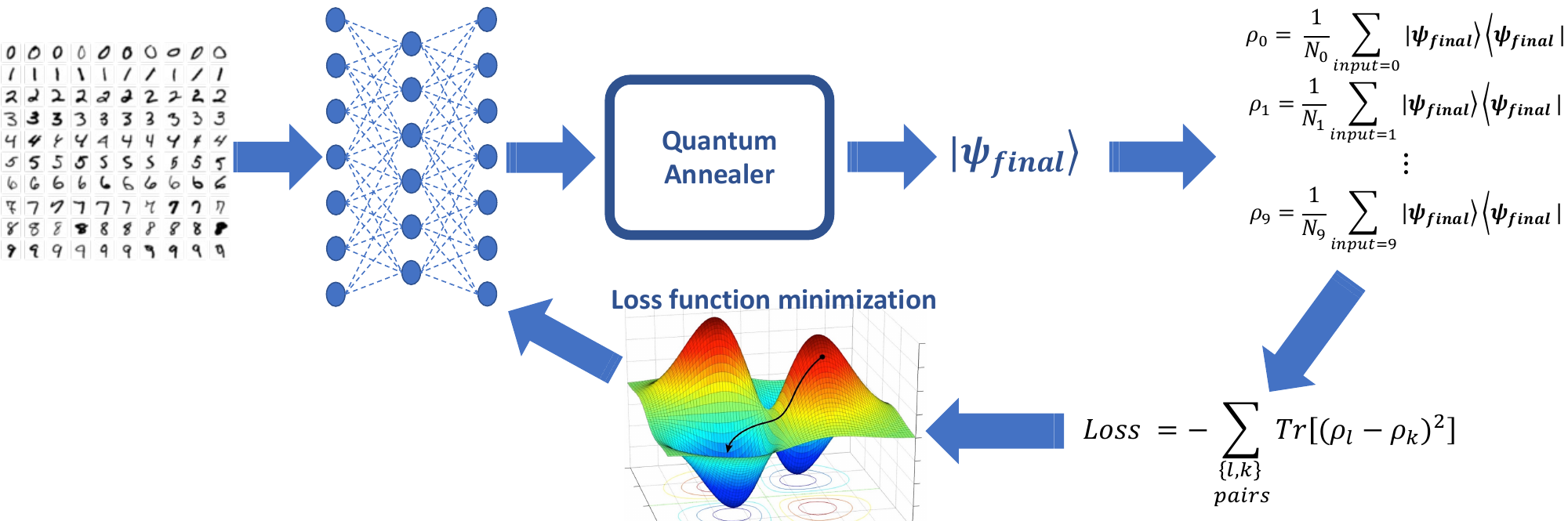}
    \caption{
    \emph{Training procedure.} Training the model works as follows. First, the neural network receives (classical) input data and feeds it to the quantum annealer, which outputs a state $\ket{\psi_{final}}$ for each input. For details, see figure (\ref{fig:setup}) and section \ref{sec:setup_and_simulations}. In the above example, the inputs are $28$ by $28$ grayscale MNIST images, corresponding to hand-written digits from $0$ to $9$. Second, all the $\ket{\psi_{final}}$ corresponding to the same class are grouped, and a density matrix is calculated for each class, as shown in the top right corner. Third, the Hilbert-Schmidt distance is calculated between all pairs of density matrices. Intuitively, this can be understood as the average distance between all the clusters of states on a Bloch sphere, as shown in a $1$-qubit example in figure (\ref{fig:Bloch}). Forth, the loss function of the neural network is defined as the negative Hilbert-Schmidt distance and then used to perform gradient descent on the neural network's parameters. These four steps are repeated until the loss function converges. Intuitively, the neural network is optimized in such a way that the output states $\ket{\psi_{final}}$ cluster into groups located as far away from each other as possible, illustrated on the right side of figure (\ref{fig:Bloch}). Once the neural network is trained well, we can use it to perform classification on new input data because the input will be mapped into a quantum state close to its corresponding cluster. 
    }
    \label{fig:training}
\end{figure*}

This work investigates a hybrid quantum-classical approach to performing classification tasks. The setup is illustrated and described in figures (\ref{fig:training}) and (\ref{fig:setup}).In this study, we consider a classical fully-fledged neural network is connected with a small quantum annealer. In such a setting, the fully-fledged classical neural-network already has built-in nonlinearity and can already handle the classification problem alone, we want to see whether an additional quantum layer could boost its performance. Supervised training is performed on classical labeled datasets, such as MNIST and CIFAR, shown in figure (\ref{fig:MNIST_zero_and_one}) and (\ref{fig:MNIST_and_CIFAR}). The raw data is fed into a (classical) neural network. But instead of performing the entire learning procedure with it (i.e., directly reading the outputs of the neural network for classification), we feed the neural network's output as the input to a quantum annealer. The quantum annealer interprets the neural network's output as the annealing schedule, as described in section (\ref{sec:setup_and_simulations}). 
Once the quantum annealer in figures (\ref{fig:training}) and (\ref{fig:setup}) outputs a final quantum state $\ket{\psi_{final}}$, this is then used to calculate the loss function, which is then used to optimize the NN parameters via gradient descent. The complete training procedure is shown in figure (\ref{fig:training}).

\section{Related work}
\subsection{Quantum Machine Learning}
As quantum technologies continue to advance rapidly, it becomes increasingly important to understand which applications can benefit from the power of these devices. Over the past decades, ML on classical computers has made significant progress. Advanced ML and artificial intelligence (AI) tools have become openly available and easily accessible through libraries such as TensorFlow \cite{abadi2016tensorflow}, Scikit-learn \cite{pedregosa2011scikit}, and PyTorch \cite{paszke2019pytorch} – to name a few. Together with increasing computing power, this resulted in revolutionizing applications such as text and image recognition, translation, and recently even the development of models for generating digital images from natural language descriptions – such as \textit{DALL-E 2} \cite{marcus2022very}. Nevertheless, progress in applications such as self-driving cars \cite{badue2021self} has been slower than many would have hoped, reminding us of the limitations of ML and AI. Therefore, if quantum computers could accelerate ML, the potential for impact would be significant. 

Two major paths to the quantum enhancement of ML have been considered \cite{huang2021power}. First, driven by quantum applications in optimization \cite{farhi2001quantum, durr1996quantum, rebentrost2014quantum, harrow2009quantum, grover1996fast}, the power of quantum computing could be used to help improve the learning process of existing classical models \cite{neven2009training}, or improve inference in graphical models \cite{leifer2008quantum}. This could result in better methods for finding optima in a training landscape or with fewer queries. However, in most general unstructured problems, the advantage of these algorithms compared to their classical counterparts may be limited to quadratic or small polynomial speedups \cite{mcclean2021low, aaronson2009need}.

The second area of interest is the possibility of using quantum models to express functions that are inefficient to represent by classical computation \cite{huang2021power}. Recent theoretical and experimental successes demonstrating quantum computations beyond classical tractability can be considered evidence that quantum computers can sample from probability distributions that are exponentially difficult to sample from classically \cite{arute2019quantum, boixo2019quantum}. It is important to emphasize that these distributions have no known real-world applications. This is the type of advantage sought in recent work on quantum neural networks (QNNs) \cite{mcclean2016theory, farhi2018classification, peruzzo2014variational}, which seek to parameterize a distribution through a set of adjustable parameters. Nevertheless, it is well known that these QNNs come with a host of trainability issues that must be addressed before they are implemented in large-scale systems \cite{mcclean2018barren,cerezo2021cost,anschuetz2022quantum}.

\begin{figure*}[htp]
    \centering
    \includegraphics[width=0.9\textwidth]{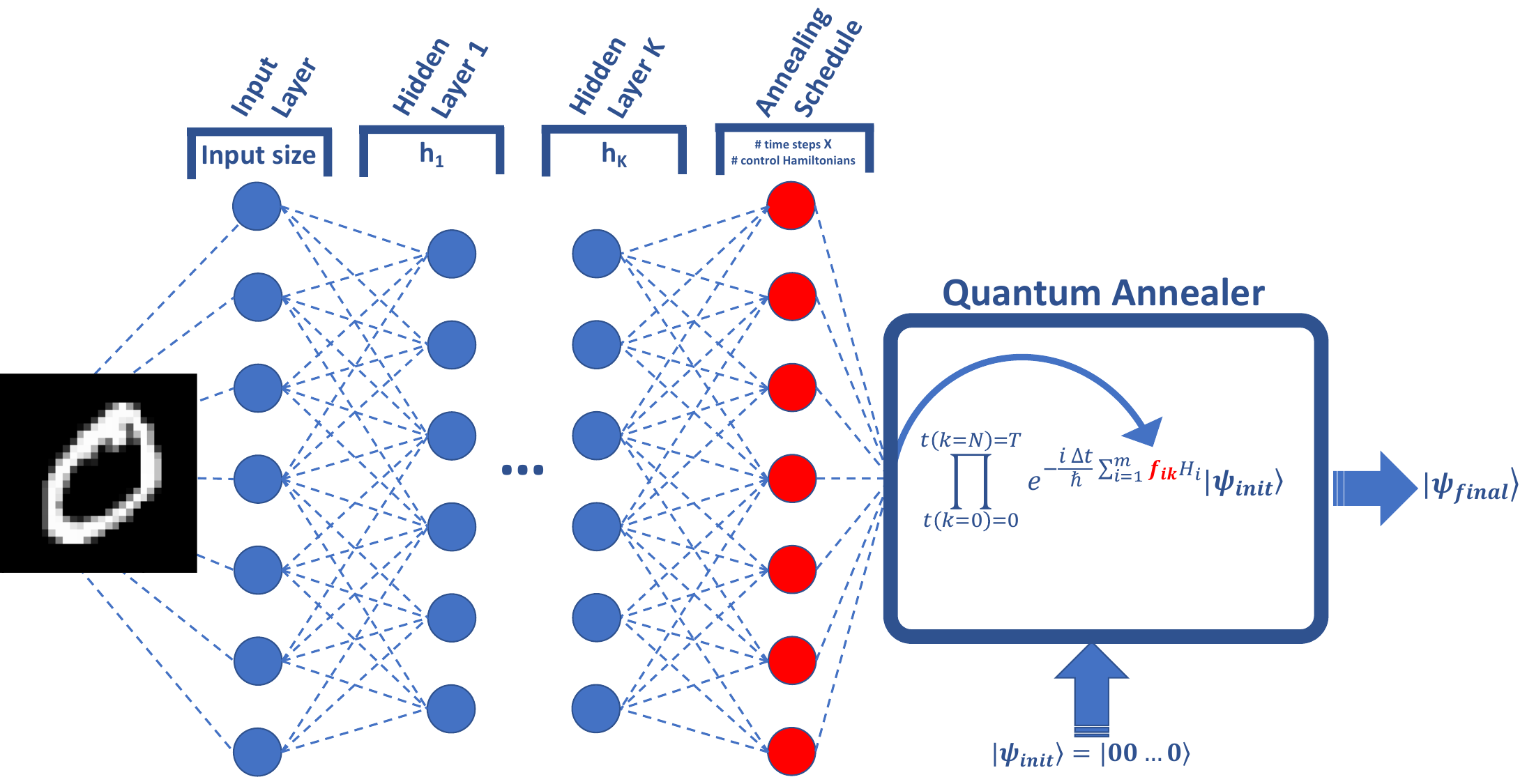}
    \caption{
    Hybrid classical neural network/quantum annealer setup. The setup works as follows: the neural network receives classical data as input and forward-passes the information through its layers. In the above case, it's a $28$ by $28$ grayscale MNIST image, and the neural network has only one hidden layer with $120$ neurons. The outputs of the neural network, \textcolor{red}{$f_{ik}$}, are the annealing schedule for the quantum annealer. The quantum annealer starts with state $\ket{0,...,0}$ and outputs state $\ket{\psi_{final}}$, which is defined above. The value of \textcolor{red}{$i$} corresponds to the different Hamiltonians which the quantum annealer can apply (see the beginning of section \ref{sec:setup_and_simulations}), and the value of \textcolor{red}{$k$} corresponds to the discretized time instance of the annealing schedule. The output state $\ket{\psi_{final}}$ can then be used for training the model, as shown in figure (\ref{fig:training}), as well as performing classification tasks. 
    }
    \label{fig:setup}
\end{figure*}

\subsection{Kernel methods}
Some quantum algorithms, such as the quantum principal component analysis \cite{lloyd2014quantum}, and quantum support vector machines \cite{rebentrost2014quantum}, have been inspired by classical machine learning algorithms. This field is generally known as \textit{quantum machine learning (QML)} \cite{biamonte2017quantum, wittek2014quantum}. Similar approaches have been used to create quantum algorithms inspired by classical neural networks. 

It has been shown that the mathematical frameworks of many quantum machine learning (QML) models are similar to \textit{kernel methods} \cite{schuld2021supervised}. Quantum computers and kernel methods process information by mapping it into high (or infinite) dimensional vector spaces without explicitly calculating each element of this vector space explicitly. In this work, we will focus on QML in the context of learning classical data, even though QML can be applied to quantum data as well \cite{aimeur2006machine, schuld2018supervised}. To perform computation, data must be encoded into the quantum system's physical states, equivalent to a \textit{feature map} \cite{schuld2019quantum, havlivcek2019supervised}. This map assigns quantum states to classical data, and the inner product of the encoded quantum states gives rise to a kernel \cite{havlivcek2019supervised, schuld2020circuit, chatterjee2017generalized, rebentrost2014quantum}. 
In classical kernel methods, we access the feature space through kernels or inner products of feature vectors. In quantum kernel methods, we also indirectly access the high-dimensional feature space through inner products of quantum states, which are facilitated by measurements. While the Hilbert space of quantum states may be vast, deterministic quantum models can still be trained and operated in a lower-dimensional subspace. Thus, we obtain the benefits of kernel methods without requiring direct access to the entire Hilbert space.
The equivalence of QML and kernel methods resulted in research on kernelized ML models \cite{blank2020quantum}, quantum generative models \cite{liu2018differentiable}, as well as understanding the difference between the complexities of quantum and classical ML algorithms \cite{havlivcek2019supervised, liu2021rigorous, huang2021power} – to mention a few examples. 


The connection between kernel and quantum learning methods arises from the classical data encoding process (quantum embedding), which determines the kernel. Although the quantum kernel has the potential to access an extremely high-dimensional Hilbert space, quantum learning model training actually occurs in a lower-dimensional subspace. This contrasts with variational models, where optimizing a significant number of classical parameters and finding a suitable circuit ansatz are necessary. The ansatz's parameters also require optimization. It's important to note that quantum learning models, including quantum neural networks (QNNs), do not always require exponentially many parameters to achieve effective results.


It is important to consider the limitations of kernel methods, as highlighted by Kübler et al.\cite{kubler2021inductive}. The authors suggest that quantum machine learning models can only offer speed-ups if we manage to encode knowledge about the problem at hand into quantum circuits, while encoding the same bias into a classical model would be hard. They argue that this situation may plausibly occur when learning on data generated by a quantum process, but it seems more challenging for classical datasets \cite{kubler2021inductive}. Furthermore, the authors show that finding suitable quantum kernels is not easy because the kernel evaluation might require exponentially many measurements, which could limit the practicality of quantum kernel methods in some cases \cite{kubler2021inductive}. Despite the potential benefits of quantum kernel methods, these limitations should be taken into account when considering their use in quantum machine learning.

\begin{figure}
    \centering
    \begin{subfigure}{0.5\textwidth}
   \centering
   \includegraphics[width=1.0\textwidth]{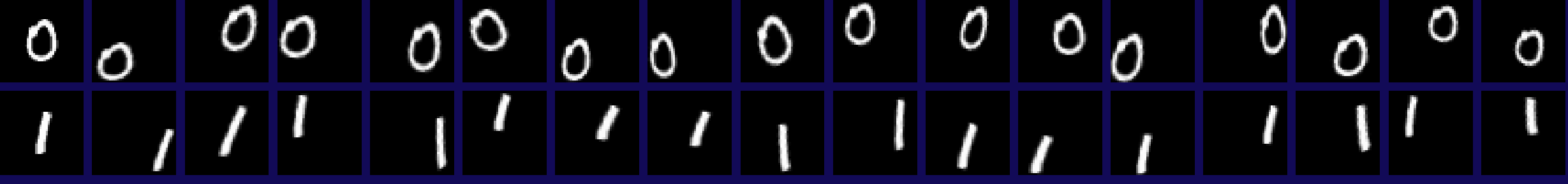} 
   \caption{
   MNIST digits. In this toy model, we use $28$ by $28$ grayscale MNIST images corresponding to $0$s and $1$s.
   }
   \label{fig:MNIST_zero_and_one}
    \end{subfigure}\hfill

    \begin{subfigure}{0.5\textwidth}
   \centering
   \includegraphics[width=1.0\textwidth]{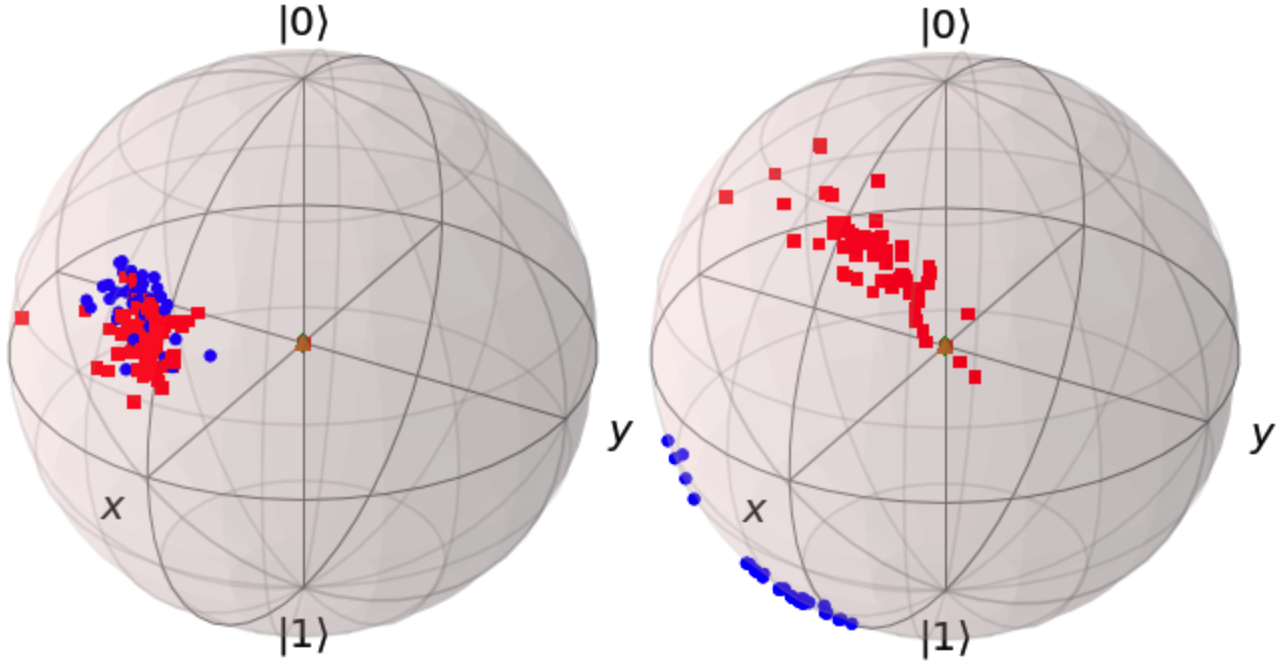} 
   \caption{
   Quantum annealer output states of a toy model. The Bloch sphere on the left side shows the output states of a $1$-qubit quantum annealer setup shown in figure (\ref{fig:setup}), \textbf{before} any training has been performed. The neural network has been randomly initialized, so there is no way of telling apart the states corresponding to $0$ input and $1$ input (shown in blue and red, respectively). The Bloch sphere on the right side shows the output states of the same quantum annealer setup, but \textbf{after} the neural network parameters have been optimized through the training procedure shown in figure (\ref{fig:training}). Now, all the output states are mapped into two visibly distinct clusters. The same principle also applies to classification with $K>2$ classes in $n>1$ qubits, except we train the model by maximizing the average distance between the clusters.
   }
   \label{fig:Bloch}
    \end{subfigure}
    \caption{
    A simplified example of training our hybrid neural network/quantum annealer from figure (\ref{fig:setup}) through the training procedure shown in figure (\ref{fig:training}).
    }
    \label{fig:one_qubit_example_MNIST}
\end{figure}

\subsection{Quantum Annealing}
An in-depth explanation can be found in Albash \textit{et al.}, 2018 \cite{albash2018adiabatic}, but for our purposes, it suffices to define adiabatic quantum annealing as a computational model in which the system remains in an instantaneous ground state of a time-dependent Hamiltonian $H(t)$ \cite{crosson2021prospects}. For example, define 
\begin{equation}\label{eq:hamiltonian_general}
    H(t) = A(t) H_X + B(t) H_Z
\end{equation}
where $H_X = - \sum_i \sigma_X^{(i)}$ denotes the Pauli $\sigma_x$ matrices acting on the $i$-th qubit and $H_Z$ is a Hamiltonian that is diagonal in the computational basis of eigenstates of tensor products of $\sigma_Z^{(i)}$.$A(t)$ and $B(t)$ are the time-dependent transverse and longitudinal coupling strengths. $H_Z$ is the problem Hamiltonian, and its ground states encode the solution to a computational problem. In the case of standard quantum annealing, we start with $A(t=0)=A_0$ and then monotonically decrease this to $A(t=T)=0$ during a time interval $[0,T]$. During the same time period, $B(t)$, which is the control of our problem Hamiltonian, $H_Z$, is monotonically increased from $B(t=0)=0$ to $B(t=T)=B_0$. The core assumption here is that this transition happens slowly enough so that the system (which started in the ground state of $H_X$) remains in the ground state of $H(t)$ until it eventually ends up in the ground state of $H_Z$ at time $t=T$. This gives us access to the solution to our computational problem. In a closed system, the time evolution of the entire system can be described by
\begin{equation}\label{eq:time_evolution}
    \ket{\psi_{final}} = \Big( \int_0^T e^{-\frac{i t}{\hbar} H(t)} dt \Big) \ket{\psi_{initial}}
\end{equation}

In this work, we are not just focused on the adiabatic limit of quantum annealing (where the system always remains in the ground state of $H(t)$) but also in the diabatic setting. Here, diabatic transitions from and to low-energy excited states are permitted. The term \textit{diabatic quantum annealing} first appeared in \textit{Muthukrishnan et al., 2016} \cite{muthukrishnan2016tunneling}, although the same concept appeared in earlier literature as \textit{nonadiabatic quantum annealing} \cite{katsuda2013nonadiabatic}, and later also as \textit{pulsed quantum annealing} \cite{karanikolas2020pulsed}.\\

With this in mind, we can now define \textit{diabatic quantum computing}. To do so, we relax the condition that the system must always remain in a single instantaneous eigenstate of $H(t)$. Instead, diabatic quantum computing and annealing are computational models (universal and for optimization, respectively) in which the system remains in a subspace spanned by the eigenstates of $H(t)$ at all times. 

The instantaneous computational eigenstate in quantum annealing and adiabatic quantum computing is usually taken to be the ground state of $H(t)$. Therefore, the energy band in diabatic quantum computing is usually taken to be the low-energy subspace of $H(t)$, even though, by definition, it could be any part of the energy spectrum. \\

It is important to point out that in diabatic quantum computing, the final state reached in equation (\ref{eq:time_evolution}) doesn't have to be the ground state of $H(t)$. 
Diabatic quantum annealing dynamics are generally non-local because the unitary term in equation (\ref{eq:time_evolution}) cannot be written as a tensor product of locally acting unitaries. This is due to the non-commutativity of the terms in equation (\ref{eq:hamiltonian_general}). The standard method for digitalizing this non-local unitary on a gate-based device required a very large number of gates \cite{haah2021quantum}, meaning that diabatic quantum annealing hardware can be used to run algorithms distinct from those being implemented on gate-based devices. 


\section{Setup and Simulations}\label{sec:setup_and_simulations}

To illustrate the idea behind the training procedure, we present a simplified example, illustrated in figure (\ref{fig:one_qubit_example_MNIST}). We use a subset of the MNIST database of handwritten digits, containing only $0$s and $1$s – shown in figure (\ref{fig:MNIST_zero_and_one}). And we use our training setup from figure (\ref{fig:training}) with a small NN to train our model in figure (\ref{fig:setup}). For simplicity, we use $n=1$ qubits, so we can visualize all states on Bloch spheres. The NN is randomly initialized, so after passing the input data through the small NN and then inputting the NN's output into the schedule parameters of the quantum annealer to evolve its initial state into a final state, we will end up with a cluster of randomly mixed $\psi_{output}$ states. This pre-training stage is illustrated on the left Bloch sphere in figure (\ref{fig:Bloch}). The loss function, which is the negative average Hilbert-Schmidt distance, defined in equation (\ref{eq:hilber-schmidt}), is close to zero. Next, we start training the NN by minimizing the loss function (or maximizing the  Hilbert Schmidt distance between states corresponding to different classes), until we eventually end up with a trained NN. Now, if we pass the same (or new) images of $0$s and $1$s through the trained NN and then pass the NN's output to the quantum annealer to get evolved final states, we end up with states as shown on the right of figure (\ref{fig:Bloch}). Here, we can clearly see that the input data was mapped into two distinct clusters, corresponding to $0$s and $1$s, respectively.

The hardware we simulated is able to apply the following operators on the state inside the quantum computers:
\begin{itemize}\label{list:Hamiltonians}
    \item $\sigma_x$ operation on individual qubits: $\begin{pmatrix}
  0&1\\
  1&0
    \end{pmatrix}$
    \item $\sigma_z$ operation on individual qubits: $\begin{pmatrix}
  1&0\\
  0&-1
    \end{pmatrix}$
    \item  $\sigma_z \otimes \sigma_z$ on each pair of qubits
\end{itemize}

The hardware has the ability to apply any linear combination of these operators at a given time and can vary their individual intensities over time. In total, for an $n$-qubit quantum system, there are $m = n + n + \frac{n(n-1)}{2} = \frac{n}{2}(n+3)$ Hamiltonians, and therefore also $m$ parameters which control the system at any instance of time.

Our goal is to create a hybrid NN, which consists of a classical NN and a quantum diabatic annealer. The Hamiltonian at any time instance can be described as a linear combination of the form:

\begin{equation}\label{eq:Hamiltonian_linear_combination}
    H(t) = \sum_{i=1}^{m}f_i(t)H_i
\end{equation}

where $H_i$ for $i \in \{1,...,n\}$ refers to single $\sigma_x$ gates acting on the $i^{th}$ qubit. $H_i$ for $i \in \{n+1,...,2n\}$ refers to single $\sigma_z$ gates acting on the $(i-n)^{th}$ qubit. $H_i$ for $i \in \{2n+1,...,m\}$ refers to double $\sigma_z \otimes \sigma_z$ gates acting on all pairs of $(i,j) \in \{1,...,n\}$. The functions $f_i(t)$ (or \textcolor{red}{$f_{ik}$} in the discretized version from figure \ref{fig:setup}) define "how much" of the Hamiltonian $H_i$ is applied at the time instance $t$. Here $t \in [0,T]$, meaning that the Hamiltonians are applied within a fixed time period $T$.

The goal of this simulation is to find the right weights and biases of the classical NN which, after classical input data is forward passed through it, outputs the optimal $f_i(t)$ functions which map the initial quantum state $\ket{\psi_{init}}=\ket{0...0}$ to \textit{maximally} distant final quantum states $\ket{\psi_{final}}$ for input data corresponding to distinct classes. At the same time, for input data corresponding to the same class, it should map the initial quantum state $\ket{\psi_{init}}$ to \textit{minimally} distant final quantum states $\ket{\psi_{final}^{(k)}}$. Here $k$ is simply the label of a class for classification purposes. In other words, $\ket{\psi_{final}^{(k)}}$ states form clusters, as shown in a $n=1$ qubit example with two classes in figure (\ref{fig:one_qubit_example_MNIST}). \\

Let us explain the details of the whole setup, shown in figure \ref{fig:setup}, starting with the classical NN on the left. As aforementioned, the input to the NN can be any type of labeled classical dataset (not restricted to images). We used different types of NNs, depending on how complex a given dataset is. For MNIST, shown in figure (\ref{fig:MNIST_and_CIFAR}), a simple NN with one hidden layer with $120$ neurons was sufficient. More complex datasets, such as CIFAR \cite{krizhevsky2009learning}, also shown in figure (\ref{fig:MNIST_and_CIFAR}), required convolutional neural networks (CNNs) with several hidden layers. It is common practice in the ML community to normalize all the input data $D$, so that $max{(D)}=1$, $min{(D)}=-1$, and $mean{(D)}=0$. The weights and biases of the NN are initialized randomly. 

For example, in the case of MNIST (as shown in figure \ref{fig:setup}.), using $\operatorname{tanh}$ activation functions for connecting the input layer with the hidden layer, followed by $Sigmoid$ activation functions connecting the hidden layer with the output layer worked well. Once the input data has been passed through the randomly initialized NN, we receive the annealing schedule, which is the input to the quantum annealer shown in figure (\ref{fig:setup}). As the output of the NN is supposed to describe the annealing schedules $f_i(t)$, we have to restrict $f_i(t)$ to a set of discrete values, as opposed to continuously changing functions. Therefore, we have to choose a value for the total number of discrete "steps" that each of the functions $f_i(t)$ can take. Hence, the actual output of the NN isn't continuous functions $f_i(t)$, but rather discrete values \textcolor{red}{$f_{ik}$}, as shown in figure (\ref{fig:setup}). The total size of the NN's output layer is the number of discretized steps multiplied by the number of control Hamiltonians:\\

NN output size = $steps$ $\times$ $m$=$steps$ $\times$ $\frac{n}{2}(n+3)$ \\

Next, let us talk about the quantum annealer, shown in the right half of figure \ref{fig:setup}. The quantum annealer is initialized with the standard $\ket{\psi_{input}}=\ket{0,...,0}$ quantum state (which is the ground state of a pure longitudinal Hamiltonian without interactions), and then evolves this state using the Hamiltonian $H(t)$ described in equation (\ref{eq:Hamiltonian_linear_combination}). Given that the Hamiltonian $H(t)$ in our discretized model changes only every $\Delta t = \frac{T}{steps}$, the change of the quantum state inside the quantum annealer between each discretized time instance can be described by:

\begin{equation}\label{eq:matrix_exponetiation}
    \ket{\psi({t_{k+1}})} = e^{\frac{-i \Delta t}{\hslash} H(t_k)} \ket{\psi(t_k)} 
\end{equation}

where $t_k = k\Delta t$ and $H(t) = \sum_{i=1}^{m}f_i(t)H_i$.

The final quantum state is:

\begin{equation}\label{eq:matrix_exponetiation_enitre_time}
    \ket{\psi_{final}} = \prod_{t(i=0)=0}^{t(i=N)=T} e^{-\frac{i \Delta t}{\hslash} \sum_{i=1}^{m} \textcolor{red}{f_{ik}} H_i } \ket{0,...,0}
\end{equation}

where $f_{ik} = f_i(t_k)$. Therefore, given all parameters of the classical NN, any input to the entire setup in figure (\ref{fig:setup}) is mapped to a particular quantum state. \\

\subsection{Training procedure}
In the previous section, we described how a classical input state (for example, a handwritten digit) can be mapped into a quantum state through the setup shown in figure (\ref{fig:setup}). In this section, we will explain how this setup can be used for classification tasks. A key element of every ML application is a training procedure, which uses a training dataset. The whole training procedure is illustrated in figure (\ref{fig:training}).

The first part of the training procedure is just repeating the steps described in section \ref{sec:setup_and_simulations}. We take the whole dataset of, in this illustration, MNIST $28$ by $28$ images, pass them through the NN, and then use the outputs of the NN as annealing schedules to create a set of output states $\ket{\psi_{final}}$. As a next step, we create a set of mixed-density matrices for every class in the dataset. For a standard MNIST dataset, there are $K=10$ classes. One for each digit. The density matrix corresponding to class $j$ is defined as $\rho_j = \sum_{i \in \{\text{input}_i \text{ = j}\}} \ket{\psi_{final}}\bra{\psi_{final}}$. The goal of the training procedure is to make these density matrices as distinct as possible. We used the Hilbert-Schmidt distance as a metric to compare two density matrices. The Hilbert-Schmidt distance is defined as:
\begin{equation}\label{eq:hilber-schmidt}
    D_{HS}(\rho,\sigma) = Tr\big[(\rho-\sigma)^2\big] = ||\rho - \sigma||_2^2
\end{equation}

We could have also used the standard trace distance, defined as $D(\rho,\sigma) = \frac{1}{2}Tr\big[\rho - \sigma\big] = \frac{1}{2}||\rho - \sigma||_1$. We decided not to use it, as the trace distance is more difficult to compute than the Hilbert-Schmidt distance, as it involves the diagonalization of $\rho - \sigma$ \cite{coles2019strong}.

Even though a mixed-density matrix cannot be shown on a Bloch sphere, such as in figure (\ref{fig:Bloch}), we can still imagine the training procedure to be somewhat similar. Our goal is to maximize the distance between states corresponding to different classes while keeping the distance between states corresponding to the same class as small as possible. Therefore, we define the loss function of the NN as minus the average Hilbert-Schmidt distance between all  pairs of density matrices $\rho_j$. Having a loss function allows us to use gradient descent (or another minimization algorithm) on all the parameters of the NN. This gradient-update procedure now needs to be repeated until the loss function converges to a minimum (=maximum average distance).

\subsection{Computational methods and errors}
In setting up the hybrid NN-QA model in this study, the choice was made to work with the largest number of qubits possible, as determined by practical memory and computation time limits on a conventional computer. The main bottleneck for the runtime was calculating the matrix exponentiations. The main bottleneck for memory consumption was performing the backpropagation (gradient calculation) step during the NN training procedure. Also, before even looking into the results of our simulations, it is essential to determine the magnitudes of the error rates of our approximation methods and whether or not they accurately simulate the quantum annealer. All numerical methods used are already well-established, including the matrix exponentiation approximation method needed for simulating equation (\ref{eq:matrix_exponetiation}). This method is based on the SSA approximation \cite{al2010new,arioli1996pade}, and it is one of the fastest and most accurate methods for general, non-sparse matrix exponentiation. Nevertheless, the computational resources needed for this matrix exponentiation approximation scales roughly as $\mathcal{O}(2^n)^3$. Therefore, given the large number of matrix exponentiations used in our simulations, we had to implement the Suzuki-Trotter approximation, explained in more detail in appendix \ref{sec:trotter}. The Suzuki-Trotter approximation method can be made arbitrarily accurate by adjusting its parameters, resulting in a trade-off between accuracy and the computational resources needed. A more detailed investigation of this method can be found in appendix \ref{sec:trotter}.

\section{Results}\label{sec:results}
In this section we discuss the datasets used,  the classification accuracies achieved with them, as well as the overall improvement through the quantum annealer.

\subsection{Datasets and learning models}
\begin{figure}
    \centering
    \begin{subfigure}{0.23\textwidth}
   \centering
   \includegraphics[width=0.85\textwidth]{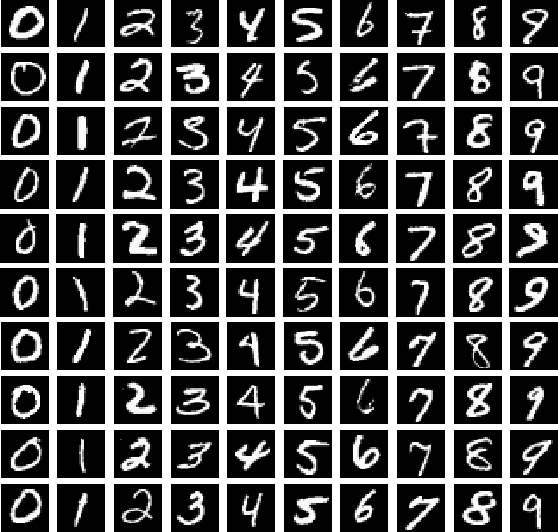} 
    \end{subfigure}\hfill
    \begin{subfigure}{0.23\textwidth}
   \centering
   \includegraphics[width=1.06\textwidth]{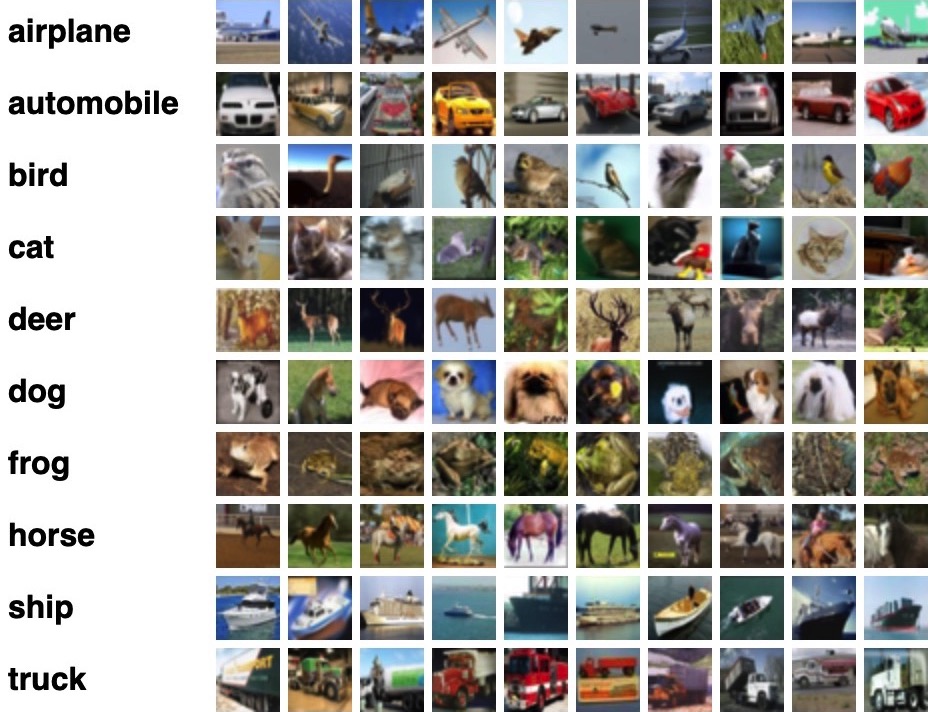} 
   \label{fig:CIFAR_dataset}
    \end{subfigure}
    \caption{a) MNIST dataset from \cite{lecun1998mnist}. It consists of $28$ by $28$ grayscale images ($784$ numbers each between $0$ and $127$). The full training dataset contains 60,000 images and the testing dataset 10,000 images.\newline b) CIFAR dataset from \cite{krizhevsky2009learning}. It consists of $32$ by $32$ color images ($1024$ triples of numbers, each between $0$ and $127$). The full training dataset contains 50,000 images, and the testing dataset 10,000 images. }
    \label{fig:MNIST_and_CIFAR}
\end{figure}

In this section, we analyze the classification accuracies of four different setups on four different datasets. The datasets listed in table \ref{tab:datasets} are MNIST, CIFAR, ISOLET, and UCIHAR:
\begin{itemize}
    \item \textbf{MNIST} consists of $28$ by $28$ grayscale images of handwritten digits from $0$ to $9$, shown in figure (\ref{fig:MNIST_and_CIFAR}). As of writing, the highest classification accuracy is $99.91\%$ achieved by using a majority vote between several convolutional neural networks \cite{an2020ensemble}. 
    \item \textbf{CIFAR} consists of $32$ by $32$ color images of $10$ different objects, shown in figure (\ref{fig:MNIST_and_CIFAR}). As of writing, the highest classification accuracy achieved is $99.5\%$ by using transformer models, shaking up the long "supremacy" of CNNs \cite{dosovitskiy2020image}.
    \item \textbf{ISOLET} consists of human voice recordings pronouncing the letters of the English alphabet, hence $26$ different classes
    \item Lastly, \textbf{UCIHAR} consists of acceleration and velocity data from a smartphone's gyroscope, as well as other motion, time, and frequency variables. The $12$ labeled classes include activities such as walking upstairs, sitting, standing, etc. 
\end{itemize}

\begin{table}[h]
\centering
\caption{Datasets ($n$: \# of features, $k$: \# of classes). 
}
\label{tab:datasets}
\resizebox{0.5\textwidth}{!}{
\begin{tabular}{ccclcccc}
\hline
\hline
& $n$ & $K$ & \shortstack{\textbf{Data}\\ \textbf{Size}} & \shortstack{\textbf{Train}\\ \textbf{Size}} & \shortstack{\textbf{Test}\\ \textbf{Size}} & \textbf{Description}  & Source  \\ \hline
\textbf{MNIST} & 784 & 10 & 11.1MB  & 60,000 & 10,000 & Image recognition of handwritten digits (fig. \ref{fig:MNIST_and_CIFAR}) & \cite{lecun1998mnist}   \\ 
\textbf{CIFAR} & 3072 & 10 & 163MB  & 50,000 & 10,000 & Image recognition of different objects (fig. \ref{fig:MNIST_and_CIFAR})  & \cite{krizhevsky2009learning} \\ 
\textbf{ISOLET} & 617 & 26 & 19MB   & 6,238 & 1,559 & Isolated letter speech recognition & \cite{smith2001speech} \\ 
\textbf{UCIHAR} & 561 & 12 & 10MB   & 6,213  & 1,554 & Human activity recognition with smartphones & \cite{anguita2013public}
\\ 

\hline
\hline
\end{tabular}
}
\end{table}

\begin{table*}[htp]
\caption{Classification accuracies and numbers of trainable parameters. This is for n=10 qubits, steps=10, TN=50. These results are discussed in section (\ref{sec:quantum_learning}). The reason why our (C)NN's do not have the same number of parameters as the (C)NN's + Annealer's is that the output layer is of different sizes. The (C)NN has $K$ output features (one for every class), while the (C)NN+Annealer has one output for every parameter of the annealing schedule. }

\centerline{
\resizebox{0.98\textwidth}{!}{
\begin{tabular}{cccccccccccccc}
\toprule
\textbf{Architecture} & \textbf{\begin{tabular}[c]{@{}c@{}}Training\end{tabular}} & \multicolumn{2}{c}{\textbf{MNIST 10}} & \multicolumn{2}{c}{\textbf{CIFAR 10}}   & \multicolumn{2}{c}{\textbf{ISOLET 26}}  & \multicolumn{2}{c}{\textbf{UCIHAR 12}} \\ 
    &   & Accuracy    & Params    & Accuracy   & Params   & Accuracy   & Params   & Accuracy  &  Params \\
\hline
\textbf{NN}   & No  & $(10.0 \pm 0.2)\%$ & $97,346$  & $(10.0 \pm 0.2)\%$  & $371,906$   & $(3.8 \pm 0.2)\%$  & $77,306$   & $(8.3 \pm 0.5)\%$  & $70,586$  \\ 
\textbf{NN}  & Yes  & $(88.2 \pm 0.3)\%$ & $97,346$  & $(44.1 \pm 0.4)\%$   & $371,906$   & $(86.2 \pm 0.3)\%$  & $77,306$   & $(87.2 \pm 0.2)\%$  & $70,586$  \\ 
\textbf{NN + Annealer} & No & $(59 \pm 1)\%$ & $133,525$  & $(12 \pm 2)\%$  & $408,085$    & $(43.9 \pm 0.4)\%$ & $113,485$   & $(44.2 \pm 0.3)\%$  & $106,765$  \\
\textbf{NN + Annealer} & Yes & $(86 \pm 1)\%$ & $133,525$  & $(13 \pm 2)\%$ & $408,085$   & $(85.5 \pm 0.4)\%$ & $113,485$  & $(85.9 \pm 0.4)\%$  & $106,765$ \\
\textbf{CNN}   & No  & $(10.0 \pm 0.2)\%$ & $45,786$ & $(10.0 \pm 0.2)\%$  & $63,366$  & $(3.9 \pm 0.1)\%$  & $63,366$  & $(8.3 \pm 0.5)\%$  & $63,366$ \\ 
\textbf{CNN}  & Yes  & $(95.0 \pm 0.3)\%$ & $45,786$ & $(49.9 \pm 0.2)\%$   & $63,366$  & $(93.4 \pm 0.2)\%$  & $63,366$  & $(94.2 \pm 0.3)\%$  & $63,366$ \\ 
\textbf{CNN + Annealer} & No  & $(53 \pm 1)\%$ & $88,781$ & $(18.1 \pm 0.5)\%$ & $88,781$  & $(42 \pm 1)\%$ & $88,781$  & $(44 \pm 1)\%$  & $88,781$ \\
\textbf{CNN + Annealer} & Yes & $(91 \pm 2)\%$ & $88,781$ & $(18 \pm 1)\%$ & $88,781$  & $(90 \pm 2)\%$ & $88,781$  & $(92 \pm 1)\%$  & $88,781$ \\
\hline
\hline
\end{tabular}\label{tab:classification_accuracies}
}
}
\end{table*}

Even though the use of CNNs and transformer models can accomplish remarkable accuracies, we decided not to use such complex models. If state-of-the-art models were used, most of the learning would happen in the classical part of the model, leaving little room for improvement by the quantum annealer. Second, simulating a complex state-of-the-art model connected to a quantum annealer would be computationally infeasible, especially given the exponential growth of the quantum annealer simulation with the number of qubits.\\

Not all datasets were equally easy to learn, and each required tuning of the network architecture. Here, we either used a simple neural network with one hidden layer or a more complex convolutional neural network. We ran both of these networks in isolation and contrasted this with the setting where they were connected to quantum annealers. The four different setups used to learn the aforementioned datasets are:
\begin{itemize}
    \item \textbf{NN}: This is a very simple classical neural network consisting of an input layer with one neuron for each input feature, one hidden layer with $120$ neurons, and one output layer with one neuron for each class. The output is, therefore, $K$ real numbers between $0$ and $1$, indicating how similar the input is to each class. Training is done by classical loss function minimization. 
    \item \textbf{NN + Annealer}: This is a very simple classical neural network consisting of an input layer with one neuron for each input feature, one hidden layer with $120$ neurons, and one output layer with one neuron for each control parameter of the quantum annealer. The outputs are the parameters \textcolor{red}{$f_{ik}$} controlling the quantum annealer, as shown in figure (\ref{fig:setup}). The resulting quantum states are then used to train the neural network, as illustrated in figure (\ref{fig:training}). 
    \item \textbf{CNN}: This is a small classical convolutional neural network consisting of two 2D convolutions. The first convolution has three input channels, six output channels, and a kernel size of five, so it has $(3 \times 5 \times 5 + 1) * 6 = 456$ parameters. The second convolution has six input channels, $16$ output channels, and a kernel size of five, so it has $(6 * 5 * 5 + 1) * 16 = 2416$ parameters. An input layer with one neuron for each input feature, two hidden layers with $120$ and $84$ neurons, respectively, and an output layer with one neuron for each class. The output is, therefore, $K$ real numbers between $0$ and $1$, as shown in figure (\ref{fig:CNN}). Training is, again, done by classical loss function minimization.
    \item \textbf{CNN + Annealer}: This is a small classical convolutional neural network consisting of two 2D convolutions. The first convolution has three input channels, six output channels, and a kernel size of five, so it has $(3 * 5 * 5 + 1) * 6 = 456$ parameters. The second convolution has six input channels, $16$ output channels, and a kernel size of five, so it has $(6 * 5 * 5 + 1) * 16 = 2416$ parameters. An input layer with one neuron for each input feature, two hidden layers with $120$ and $84$ neurons, respectively, and one output layer with one neuron for each control parameter of the quantum annealer. The outputs, again, are the parameters \textcolor{red}{$f_{ik}$} controlling the quantum annealer, as shown in figure (\ref{fig:setup}). The resulting quantum states are then used to train the neural network, as illustrated in figure (\ref{fig:training}).    
\end{itemize}

\subsection{Is there any advantage in using a quantum annealer?}\label{sec:quantum_learning}


The results of our simulation experiments can be found in table \ref{tab:classification_accuracies}. We performed $32$ different experiments, with $8$ for each dataset. Firstly, it is important to note that without training, the classification accuracy of our NN and CNN was always around $\frac{1}{K}$, which is equivalent to random guessing. This is no surprise, given that the network cannot know which output neuron corresponds to which class in our prediction model. At the same time, when we connect the setup to an annealer, the classification accuracy is always better than $\frac{1}{K}$. In the case of MNIST, it is around $59\%$; for CIFAR, it is $12\%$; for ISOLET, it is $44\%$; and for UCIHAR, it is $44\%$. This is because the output vectors $\ket{\psi_{final}}$ are mapped to classes based on how close they are in the Hilbert space. Even with a neural network that is completely untrained, it is possible to achieve classification accuracies better than random choice. Nevertheless, this is not a property unique to quantum annealers but rather a consequence of the way in which classification is performed. To confirm this, we created a classical model which would average all images corresponding to a given class and then perform classification by comparing images to the $K$ different averages. For MNIST, the accuracy of this trivial model was around $63\%$ and for CIFAR around $13\%$, for ISOLET around $47\%$, and for UCIHAR also around $47\%$, with error margins of less than $1\%$. Therefore, quantum annealers provide no advantage over classical models without performing training.\\

Next, let us have a closer look at what happens after the training procedure, as shown in figure (\ref{fig:training}). The results are reported in table (\ref{tab:classification_accuracies}). We can see that training a (C)NN + Annealer model never results in better classification accuracies as compared to a classical (C)NN model. There are multiple possible explanations for this negative result:


\begin{enumerate}
    \item The annealers which we simulate may not be large enough. We could only consistently simulate systems with up to $n=10$ qubits. It is possible that with a significantly larger number of qubits an advantage is gained from using a quantum annealer
    \item It could be that we were using the wrong datasets. There is no apparent reason why, for example, image classification should benefit from a quantum annealer. All our datasets work really well with classical machine learning models, which is why they are so commonly used in research. Other datasets containing data about quantum systems might benefit from adding a quantum annealer to a neural network
    \item Lastly, it could be that there is just no benefit from using our system from figure (\ref{fig:setup}). This would mean that there  exists no quantum advantage from connecting neural networks to quantum annealers and using them to perform classical classification.
\end{enumerate}

Simulating a system with more qubits might be very difficult, so it is unlikely that significant progress can be made in this direction. Therefore, we suggest future investigations focusing on applying our system to learning inherently different datasets. An example of these could be quantum datasets, such as quantum state tomography and quantum error correction.\cite{torlai2018neural}


\section{Conclusion}
We explore a setup for performing classification on labeled classical datasets consisting of a classical neural network connected to a quantum annealer. Quantum annealers are mostly studied in the adiabatic regime, a computational model in which the quantum system remains in an instantaneous eigenstate of a time-dependent Hamiltonian. However, in our research, we focus on the diabatic regime where the quantum state doesn't always have to remain in the ground state. The neural network programs the quantum annealer's controls and thereby maps the annealer's initial states into new states in the Hilbert space. The neural network's parameters are optimized to maximize the distance of states corresponding to inputs from different classes and minimize the distance between quantum states corresponding to the same class. Recent literature \cite{yang2022analog} demonstrates that at least some of the "learning" is due to the quantum annealer in our setup. This is due to the fact that the quantum annealer introduces a non-linearity in the network. No matter how large, neural networks can only be more powerful than linear classifiers if non-linearities are introduced between layers. In \cite{yang2022analog}, the authors classified a linearly inseparable dataset by using only a linear layer + the quantum annealing setup, explicitly showing that the learning is due to the non-linearity introduced by the quantum annealer. Also, our setup can perform classification with fewer parameters for training and inference. In this study, we consider a similar but not quite the same case, where a classical fully-fledged neural network is connected with a small quantum annealer (of course, adding back the full classical nn will compromise the advantage in the number of parameters). In such a setting, the fully-fledged classical neural-network already has built-in nonlinearity and can already handle the classification problem alone, we want to see whether an additional quantum layer could boost its performance. However, as of writing, we didn't find sufficient evidence demonstrating any advantage over just using classical neural networks. This could be due to our inability to simulate large quantum annealers effectively, our datasets, or there might be no real advantage to using our setup at all.

\begin{acknowledgments}
This publication is the result of research funded by the Defense Advanced Research Projects Agency (DARPA) under agreement No.HR00112109969.

\end{acknowledgments}


\bibliography{QClassifier}

\begin{thebibliography}{67}%
\makeatletter
\providecommand \@ifxundefined [1]{%
 \@ifx{#1\undefined}
}%
\providecommand \@ifnum [1]{%
 \ifnum #1\expandafter \@firstoftwo
 \else \expandafter \@secondoftwo
 \fi
}%
\providecommand \@ifx [1]{%
 \ifx #1\expandafter \@firstoftwo
 \else \expandafter \@secondoftwo
 \fi
}%
\providecommand \natexlab [1]{#1}%
\providecommand \enquote  [1]{``#1''}%
\providecommand \bibnamefont  [1]{#1}%
\providecommand \bibfnamefont [1]{#1}%
\providecommand \citenamefont [1]{#1}%
\providecommand \href@noop [0]{\@secondoftwo}%
\providecommand \href [0]{\begingroup \@sanitize@url \@href}%
\providecommand \@href[1]{\@@startlink{#1}\@@href}%
\providecommand \@@href[1]{\endgroup#1\@@endlink}%
\providecommand \@sanitize@url [0]{\catcode `\\12\catcode `\$12\catcode
  `\&12\catcode `\#12\catcode `\^12\catcode `\_12\catcode `\%12\relax}%
\providecommand \@@startlink[1]{}%
\providecommand \@@endlink[0]{}%
\providecommand \url  [0]{\begingroup\@sanitize@url \@url }%
\providecommand \@url [1]{\endgroup\@href {#1}{\urlprefix }}%
\providecommand \urlprefix  [0]{URL }%
\providecommand \Eprint [0]{\href }%
\providecommand \doibase [0]{https://doi.org/}%
\providecommand \selectlanguage [0]{\@gobble}%
\providecommand \bibinfo  [0]{\@secondoftwo}%
\providecommand \bibfield  [0]{\@secondoftwo}%
\providecommand \translation [1]{[#1]}%
\providecommand \BibitemOpen [0]{}%
\providecommand \bibitemStop [0]{}%
\providecommand \bibitemNoStop [0]{.\EOS\space}%
\providecommand \EOS [0]{\spacefactor3000\relax}%
\providecommand \BibitemShut  [1]{\csname bibitem#1\endcsname}%
\let\auto@bib@innerbib\@empty
\bibitem [{\citenamefont {Albash}\ and\ \citenamefont
  {Lidar}(2018)}]{albash2018adiabatic}%
  \BibitemOpen
  \bibfield  {author} {\bibinfo {author} {\bibfnamefont {T.}~\bibnamefont
  {Albash}}\ and\ \bibinfo {author} {\bibfnamefont {D.~A.}\ \bibnamefont
  {Lidar}},\ }\bibfield  {title} {\bibinfo {title} {Adiabatic quantum
  computation},\ }\href@noop {} {\bibfield  {journal} {\bibinfo  {journal}
  {Reviews of Modern Physics}\ }\textbf {\bibinfo {volume} {90}},\ \bibinfo
  {pages} {015002} (\bibinfo {year} {2018})}\BibitemShut {NoStop}%
\bibitem [{\citenamefont {Kadowaki}\ and\ \citenamefont
  {Nishimori}(1998)}]{kadowaki1998quantum}%
  \BibitemOpen
  \bibfield  {author} {\bibinfo {author} {\bibfnamefont {T.}~\bibnamefont
  {Kadowaki}}\ and\ \bibinfo {author} {\bibfnamefont {H.}~\bibnamefont
  {Nishimori}},\ }\bibfield  {title} {\bibinfo {title} {Quantum annealing in
  the transverse ising model},\ }\href@noop {} {\bibfield  {journal} {\bibinfo
  {journal} {Physical Review E}\ }\textbf {\bibinfo {volume} {58}},\ \bibinfo
  {pages} {5355} (\bibinfo {year} {1998})}\BibitemShut {NoStop}%
\bibitem [{\citenamefont {Crosson}\ and\ \citenamefont
  {Lidar}(2021)}]{crosson2021prospects}%
  \BibitemOpen
  \bibfield  {author} {\bibinfo {author} {\bibfnamefont {E.}~\bibnamefont
  {Crosson}}\ and\ \bibinfo {author} {\bibfnamefont {D.}~\bibnamefont
  {Lidar}},\ }\bibfield  {title} {\bibinfo {title} {Prospects for quantum
  enhancement with diabatic quantum annealing},\ }\href@noop {} {\bibfield
  {journal} {\bibinfo  {journal} {Nature Reviews Physics}\ }\textbf {\bibinfo
  {volume} {3}},\ \bibinfo {pages} {466} (\bibinfo {year} {2021})}\BibitemShut
  {NoStop}%
\bibitem [{\citenamefont {Johnson}\ \emph {et~al.}(2011)\citenamefont
  {Johnson}, \citenamefont {Amin}, \citenamefont {Gildert}, \citenamefont
  {Lanting}, \citenamefont {Hamze}, \citenamefont {Dickson}, \citenamefont
  {Harris}, \citenamefont {Berkley}, \citenamefont {Johansson}, \citenamefont
  {Bunyk} \emph {et~al.}}]{johnson2011quantum}%
  \BibitemOpen
  \bibfield  {author} {\bibinfo {author} {\bibfnamefont {M.~W.}\ \bibnamefont
  {Johnson}}, \bibinfo {author} {\bibfnamefont {M.~H.}\ \bibnamefont {Amin}},
  \bibinfo {author} {\bibfnamefont {S.}~\bibnamefont {Gildert}}, \bibinfo
  {author} {\bibfnamefont {T.}~\bibnamefont {Lanting}}, \bibinfo {author}
  {\bibfnamefont {F.}~\bibnamefont {Hamze}}, \bibinfo {author} {\bibfnamefont
  {N.}~\bibnamefont {Dickson}}, \bibinfo {author} {\bibfnamefont
  {R.}~\bibnamefont {Harris}}, \bibinfo {author} {\bibfnamefont {A.~J.}\
  \bibnamefont {Berkley}}, \bibinfo {author} {\bibfnamefont {J.}~\bibnamefont
  {Johansson}}, \bibinfo {author} {\bibfnamefont {P.}~\bibnamefont {Bunyk}},
  \emph {et~al.},\ }\bibfield  {title} {\bibinfo {title} {Quantum annealing
  with manufactured spins},\ }\href@noop {} {\bibfield  {journal} {\bibinfo
  {journal} {Nature}\ }\textbf {\bibinfo {volume} {473}},\ \bibinfo {pages}
  {194} (\bibinfo {year} {2011})}\BibitemShut {NoStop}%
\bibitem [{\citenamefont {Harris}\ \emph {et~al.}(2010)\citenamefont {Harris},
  \citenamefont {Johansson}, \citenamefont {Berkley}, \citenamefont {Johnson},
  \citenamefont {Lanting}, \citenamefont {Han}, \citenamefont {Bunyk},
  \citenamefont {Ladizinsky}, \citenamefont {Oh}, \citenamefont {Perminov}
  \emph {et~al.}}]{harris2010experimental}%
  \BibitemOpen
  \bibfield  {author} {\bibinfo {author} {\bibfnamefont {R.}~\bibnamefont
  {Harris}}, \bibinfo {author} {\bibfnamefont {J.}~\bibnamefont {Johansson}},
  \bibinfo {author} {\bibfnamefont {A.}~\bibnamefont {Berkley}}, \bibinfo
  {author} {\bibfnamefont {M.}~\bibnamefont {Johnson}}, \bibinfo {author}
  {\bibfnamefont {T.}~\bibnamefont {Lanting}}, \bibinfo {author} {\bibfnamefont
  {S.}~\bibnamefont {Han}}, \bibinfo {author} {\bibfnamefont {P.}~\bibnamefont
  {Bunyk}}, \bibinfo {author} {\bibfnamefont {E.}~\bibnamefont {Ladizinsky}},
  \bibinfo {author} {\bibfnamefont {T.}~\bibnamefont {Oh}}, \bibinfo {author}
  {\bibfnamefont {I.}~\bibnamefont {Perminov}}, \emph {et~al.},\ }\bibfield
  {title} {\bibinfo {title} {Experimental demonstration of a robust and
  scalable flux qubit},\ }\href@noop {} {\bibfield  {journal} {\bibinfo
  {journal} {Physical Review B}\ }\textbf {\bibinfo {volume} {81}},\ \bibinfo
  {pages} {134510} (\bibinfo {year} {2010})}\BibitemShut {NoStop}%
\bibitem [{\citenamefont {Berkley}\ \emph {et~al.}(2010)\citenamefont
  {Berkley}, \citenamefont {Johnson}, \citenamefont {Bunyk}, \citenamefont
  {Harris}, \citenamefont {Johansson}, \citenamefont {Lanting}, \citenamefont
  {Ladizinsky}, \citenamefont {Tolkacheva}, \citenamefont {Amin},\ and\
  \citenamefont {Rose}}]{berkley2010scalable}%
  \BibitemOpen
  \bibfield  {author} {\bibinfo {author} {\bibfnamefont {A.}~\bibnamefont
  {Berkley}}, \bibinfo {author} {\bibfnamefont {M.}~\bibnamefont {Johnson}},
  \bibinfo {author} {\bibfnamefont {P.}~\bibnamefont {Bunyk}}, \bibinfo
  {author} {\bibfnamefont {R.}~\bibnamefont {Harris}}, \bibinfo {author}
  {\bibfnamefont {J.}~\bibnamefont {Johansson}}, \bibinfo {author}
  {\bibfnamefont {T.}~\bibnamefont {Lanting}}, \bibinfo {author} {\bibfnamefont
  {E.}~\bibnamefont {Ladizinsky}}, \bibinfo {author} {\bibfnamefont
  {E.}~\bibnamefont {Tolkacheva}}, \bibinfo {author} {\bibfnamefont
  {M.}~\bibnamefont {Amin}},\ and\ \bibinfo {author} {\bibfnamefont
  {G.}~\bibnamefont {Rose}},\ }\bibfield  {title} {\bibinfo {title} {A scalable
  readout system for a superconducting adiabatic quantum optimization system},\
  }\href@noop {} {\bibfield  {journal} {\bibinfo  {journal} {Superconductor
  Science and Technology}\ }\textbf {\bibinfo {volume} {23}},\ \bibinfo {pages}
  {105014} (\bibinfo {year} {2010})}\BibitemShut {NoStop}%
\bibitem [{\citenamefont {Berkley}\ \emph {et~al.}(2013)\citenamefont
  {Berkley}, \citenamefont {Przybysz}, \citenamefont {Lanting}, \citenamefont
  {Harris}, \citenamefont {Dickson}, \citenamefont {Altomare}, \citenamefont
  {Amin}, \citenamefont {Bunyk}, \citenamefont {Enderud}, \citenamefont
  {Hoskinson} \emph {et~al.}}]{berkley2013tunneling}%
  \BibitemOpen
  \bibfield  {author} {\bibinfo {author} {\bibfnamefont {A.}~\bibnamefont
  {Berkley}}, \bibinfo {author} {\bibfnamefont {A.}~\bibnamefont {Przybysz}},
  \bibinfo {author} {\bibfnamefont {T.}~\bibnamefont {Lanting}}, \bibinfo
  {author} {\bibfnamefont {R.}~\bibnamefont {Harris}}, \bibinfo {author}
  {\bibfnamefont {N.}~\bibnamefont {Dickson}}, \bibinfo {author} {\bibfnamefont
  {F.}~\bibnamefont {Altomare}}, \bibinfo {author} {\bibfnamefont
  {M.}~\bibnamefont {Amin}}, \bibinfo {author} {\bibfnamefont {P.}~\bibnamefont
  {Bunyk}}, \bibinfo {author} {\bibfnamefont {C.}~\bibnamefont {Enderud}},
  \bibinfo {author} {\bibfnamefont {E.}~\bibnamefont {Hoskinson}}, \emph
  {et~al.},\ }\bibfield  {title} {\bibinfo {title} {Tunneling spectroscopy
  using a probe qubit},\ }\href@noop {} {\bibfield  {journal} {\bibinfo
  {journal} {Physical Review B}\ }\textbf {\bibinfo {volume} {87}},\ \bibinfo
  {pages} {020502} (\bibinfo {year} {2013})}\BibitemShut {NoStop}%
\bibitem [{\citenamefont {Bunyk}\ \emph {et~al.}(2014)\citenamefont {Bunyk},
  \citenamefont {Hoskinson}, \citenamefont {Johnson}, \citenamefont
  {Tolkacheva}, \citenamefont {Altomare}, \citenamefont {Berkley},
  \citenamefont {Harris}, \citenamefont {Hilton}, \citenamefont {Lanting},
  \citenamefont {Przybysz} \emph {et~al.}}]{bunyk2014architectural}%
  \BibitemOpen
  \bibfield  {author} {\bibinfo {author} {\bibfnamefont {P.~I.}\ \bibnamefont
  {Bunyk}}, \bibinfo {author} {\bibfnamefont {E.~M.}\ \bibnamefont
  {Hoskinson}}, \bibinfo {author} {\bibfnamefont {M.~W.}\ \bibnamefont
  {Johnson}}, \bibinfo {author} {\bibfnamefont {E.}~\bibnamefont {Tolkacheva}},
  \bibinfo {author} {\bibfnamefont {F.}~\bibnamefont {Altomare}}, \bibinfo
  {author} {\bibfnamefont {A.~J.}\ \bibnamefont {Berkley}}, \bibinfo {author}
  {\bibfnamefont {R.}~\bibnamefont {Harris}}, \bibinfo {author} {\bibfnamefont
  {J.~P.}\ \bibnamefont {Hilton}}, \bibinfo {author} {\bibfnamefont
  {T.}~\bibnamefont {Lanting}}, \bibinfo {author} {\bibfnamefont {A.~J.}\
  \bibnamefont {Przybysz}}, \emph {et~al.},\ }\bibfield  {title} {\bibinfo
  {title} {Architectural considerations in the design of a superconducting
  quantum annealing processor},\ }\href@noop {} {\bibfield  {journal} {\bibinfo
   {journal} {IEEE Transactions on Applied Superconductivity}\ }\textbf
  {\bibinfo {volume} {24}},\ \bibinfo {pages} {1} (\bibinfo {year}
  {2014})}\BibitemShut {NoStop}%
\bibitem [{\citenamefont {Lanting}\ \emph {et~al.}(2014)\citenamefont
  {Lanting}, \citenamefont {Przybysz}, \citenamefont {Smirnov}, \citenamefont
  {Spedalieri}, \citenamefont {Amin}, \citenamefont {Berkley}, \citenamefont
  {Harris}, \citenamefont {Altomare}, \citenamefont {Boixo}, \citenamefont
  {Bunyk} \emph {et~al.}}]{lanting2014entanglement}%
  \BibitemOpen
  \bibfield  {author} {\bibinfo {author} {\bibfnamefont {T.}~\bibnamefont
  {Lanting}}, \bibinfo {author} {\bibfnamefont {A.~J.}\ \bibnamefont
  {Przybysz}}, \bibinfo {author} {\bibfnamefont {A.~Y.}\ \bibnamefont
  {Smirnov}}, \bibinfo {author} {\bibfnamefont {F.~M.}\ \bibnamefont
  {Spedalieri}}, \bibinfo {author} {\bibfnamefont {M.~H.}\ \bibnamefont
  {Amin}}, \bibinfo {author} {\bibfnamefont {A.~J.}\ \bibnamefont {Berkley}},
  \bibinfo {author} {\bibfnamefont {R.}~\bibnamefont {Harris}}, \bibinfo
  {author} {\bibfnamefont {F.}~\bibnamefont {Altomare}}, \bibinfo {author}
  {\bibfnamefont {S.}~\bibnamefont {Boixo}}, \bibinfo {author} {\bibfnamefont
  {P.}~\bibnamefont {Bunyk}}, \emph {et~al.},\ }\bibfield  {title} {\bibinfo
  {title} {Entanglement in a quantum annealing processor},\ }\href@noop {}
  {\bibfield  {journal} {\bibinfo  {journal} {Physical Review X}\ }\textbf
  {\bibinfo {volume} {4}},\ \bibinfo {pages} {021041} (\bibinfo {year}
  {2014})}\BibitemShut {NoStop}%
\bibitem [{\citenamefont {Dickson}\ \emph {et~al.}(2013)\citenamefont
  {Dickson}, \citenamefont {Johnson}, \citenamefont {Amin}, \citenamefont
  {Harris}, \citenamefont {Altomare}, \citenamefont {Berkley}, \citenamefont
  {Bunyk}, \citenamefont {Cai}, \citenamefont {Chapple}, \citenamefont {Chavez}
  \emph {et~al.}}]{dickson2013thermally}%
  \BibitemOpen
  \bibfield  {author} {\bibinfo {author} {\bibfnamefont {N.~G.}\ \bibnamefont
  {Dickson}}, \bibinfo {author} {\bibfnamefont {M.}~\bibnamefont {Johnson}},
  \bibinfo {author} {\bibfnamefont {M.}~\bibnamefont {Amin}}, \bibinfo {author}
  {\bibfnamefont {R.}~\bibnamefont {Harris}}, \bibinfo {author} {\bibfnamefont
  {F.}~\bibnamefont {Altomare}}, \bibinfo {author} {\bibfnamefont
  {A.}~\bibnamefont {Berkley}}, \bibinfo {author} {\bibfnamefont
  {P.}~\bibnamefont {Bunyk}}, \bibinfo {author} {\bibfnamefont
  {J.}~\bibnamefont {Cai}}, \bibinfo {author} {\bibfnamefont {E.}~\bibnamefont
  {Chapple}}, \bibinfo {author} {\bibfnamefont {P.}~\bibnamefont {Chavez}},
  \emph {et~al.},\ }\bibfield  {title} {\bibinfo {title} {Thermally assisted
  quantum annealing of a 16-qubit problem},\ }\href@noop {} {\bibfield
  {journal} {\bibinfo  {journal} {Nature communications}\ }\textbf {\bibinfo
  {volume} {4}},\ \bibinfo {pages} {1} (\bibinfo {year} {2013})}\BibitemShut
  {NoStop}%
\bibitem [{\citenamefont {Kaminsky}\ and\ \citenamefont
  {Lloyd}(2004)}]{kaminsky2004scalable}%
  \BibitemOpen
  \bibfield  {author} {\bibinfo {author} {\bibfnamefont {W.~M.}\ \bibnamefont
  {Kaminsky}}\ and\ \bibinfo {author} {\bibfnamefont {S.}~\bibnamefont
  {Lloyd}},\ }\bibfield  {title} {\bibinfo {title} {Scalable architecture for
  adiabatic quantum computing of np-hard problems},\ }\href@noop {} {\bibfield
  {journal} {\bibinfo  {journal} {Quantum computing and quantum bits in
  mesoscopic systems}\ ,\ \bibinfo {pages} {229}} (\bibinfo {year}
  {2004})}\BibitemShut {NoStop}%
\bibitem [{\citenamefont {Kaminsky}\ \emph {et~al.}(2004)\citenamefont
  {Kaminsky}, \citenamefont {Lloyd},\ and\ \citenamefont
  {Orlando}}]{kaminsky2004scalable2}%
  \BibitemOpen
  \bibfield  {author} {\bibinfo {author} {\bibfnamefont {W.~M.}\ \bibnamefont
  {Kaminsky}}, \bibinfo {author} {\bibfnamefont {S.}~\bibnamefont {Lloyd}},\
  and\ \bibinfo {author} {\bibfnamefont {T.~P.}\ \bibnamefont {Orlando}},\
  }\bibfield  {title} {\bibinfo {title} {Scalable superconducting architecture
  for adiabatic quantum computation},\ }\href@noop {} {\bibfield  {journal}
  {\bibinfo  {journal} {arXiv preprint quant-ph/0403090}\ } (\bibinfo {year}
  {2004})}\BibitemShut {NoStop}%
\bibitem [{\citenamefont {Abadi}\ \emph {et~al.}(2016)\citenamefont {Abadi},
  \citenamefont {Agarwal}, \citenamefont {Barham}, \citenamefont {Brevdo},
  \citenamefont {Chen}, \citenamefont {Citro}, \citenamefont {Corrado},
  \citenamefont {Davis}, \citenamefont {Dean}, \citenamefont {Devin} \emph
  {et~al.}}]{abadi2016tensorflow}%
  \BibitemOpen
  \bibfield  {author} {\bibinfo {author} {\bibfnamefont {M.}~\bibnamefont
  {Abadi}}, \bibinfo {author} {\bibfnamefont {A.}~\bibnamefont {Agarwal}},
  \bibinfo {author} {\bibfnamefont {P.}~\bibnamefont {Barham}}, \bibinfo
  {author} {\bibfnamefont {E.}~\bibnamefont {Brevdo}}, \bibinfo {author}
  {\bibfnamefont {Z.}~\bibnamefont {Chen}}, \bibinfo {author} {\bibfnamefont
  {C.}~\bibnamefont {Citro}}, \bibinfo {author} {\bibfnamefont {G.~S.}\
  \bibnamefont {Corrado}}, \bibinfo {author} {\bibfnamefont {A.}~\bibnamefont
  {Davis}}, \bibinfo {author} {\bibfnamefont {J.}~\bibnamefont {Dean}},
  \bibinfo {author} {\bibfnamefont {M.}~\bibnamefont {Devin}}, \emph {et~al.},\
  }\bibfield  {title} {\bibinfo {title} {Tensorflow: Large-scale machine
  learning on heterogeneous distributed systems},\ }\href@noop {} {\bibfield
  {journal} {\bibinfo  {journal} {arXiv preprint arXiv:1603.04467}\ } (\bibinfo
  {year} {2016})}\BibitemShut {NoStop}%
\bibitem [{\citenamefont {Pedregosa}\ \emph {et~al.}(2011)\citenamefont
  {Pedregosa}, \citenamefont {Varoquaux}, \citenamefont {Gramfort},
  \citenamefont {Michel}, \citenamefont {Thirion}, \citenamefont {Grisel},
  \citenamefont {Blondel}, \citenamefont {Prettenhofer}, \citenamefont {Weiss},
  \citenamefont {Dubourg} \emph {et~al.}}]{pedregosa2011scikit}%
  \BibitemOpen
  \bibfield  {author} {\bibinfo {author} {\bibfnamefont {F.}~\bibnamefont
  {Pedregosa}}, \bibinfo {author} {\bibfnamefont {G.}~\bibnamefont
  {Varoquaux}}, \bibinfo {author} {\bibfnamefont {A.}~\bibnamefont {Gramfort}},
  \bibinfo {author} {\bibfnamefont {V.}~\bibnamefont {Michel}}, \bibinfo
  {author} {\bibfnamefont {B.}~\bibnamefont {Thirion}}, \bibinfo {author}
  {\bibfnamefont {O.}~\bibnamefont {Grisel}}, \bibinfo {author} {\bibfnamefont
  {M.}~\bibnamefont {Blondel}}, \bibinfo {author} {\bibfnamefont
  {P.}~\bibnamefont {Prettenhofer}}, \bibinfo {author} {\bibfnamefont
  {R.}~\bibnamefont {Weiss}}, \bibinfo {author} {\bibfnamefont
  {V.}~\bibnamefont {Dubourg}}, \emph {et~al.},\ }\bibfield  {title} {\bibinfo
  {title} {Scikit-learn: Machine learning in python},\ }\href@noop {}
  {\bibfield  {journal} {\bibinfo  {journal} {the Journal of machine Learning
  research}\ }\textbf {\bibinfo {volume} {12}},\ \bibinfo {pages} {2825}
  (\bibinfo {year} {2011})}\BibitemShut {NoStop}%
\bibitem [{\citenamefont {Paszke}\ \emph {et~al.}(2019)\citenamefont {Paszke},
  \citenamefont {Gross}, \citenamefont {Massa}, \citenamefont {Lerer},
  \citenamefont {Bradbury}, \citenamefont {Chanan}, \citenamefont {Killeen},
  \citenamefont {Lin}, \citenamefont {Gimelshein}, \citenamefont {Antiga} \emph
  {et~al.}}]{paszke2019pytorch}%
  \BibitemOpen
  \bibfield  {author} {\bibinfo {author} {\bibfnamefont {A.}~\bibnamefont
  {Paszke}}, \bibinfo {author} {\bibfnamefont {S.}~\bibnamefont {Gross}},
  \bibinfo {author} {\bibfnamefont {F.}~\bibnamefont {Massa}}, \bibinfo
  {author} {\bibfnamefont {A.}~\bibnamefont {Lerer}}, \bibinfo {author}
  {\bibfnamefont {J.}~\bibnamefont {Bradbury}}, \bibinfo {author}
  {\bibfnamefont {G.}~\bibnamefont {Chanan}}, \bibinfo {author} {\bibfnamefont
  {T.}~\bibnamefont {Killeen}}, \bibinfo {author} {\bibfnamefont
  {Z.}~\bibnamefont {Lin}}, \bibinfo {author} {\bibfnamefont {N.}~\bibnamefont
  {Gimelshein}}, \bibinfo {author} {\bibfnamefont {L.}~\bibnamefont {Antiga}},
  \emph {et~al.},\ }\bibfield  {title} {\bibinfo {title} {Pytorch: An
  imperative style, high-performance deep learning library},\ }\href@noop {}
  {\bibfield  {journal} {\bibinfo  {journal} {Advances in neural information
  processing systems}\ }\textbf {\bibinfo {volume} {32}} (\bibinfo {year}
  {2019})}\BibitemShut {NoStop}%
\bibitem [{\citenamefont {Marcus}\ \emph {et~al.}(2022)\citenamefont {Marcus},
  \citenamefont {Davis},\ and\ \citenamefont {Aaronson}}]{marcus2022very}%
  \BibitemOpen
  \bibfield  {author} {\bibinfo {author} {\bibfnamefont {G.}~\bibnamefont
  {Marcus}}, \bibinfo {author} {\bibfnamefont {E.}~\bibnamefont {Davis}},\ and\
  \bibinfo {author} {\bibfnamefont {S.}~\bibnamefont {Aaronson}},\ }\bibfield
  {title} {\bibinfo {title} {A very preliminary analysis of dall-e 2},\
  }\href@noop {} {\bibfield  {journal} {\bibinfo  {journal} {arXiv preprint
  arXiv:2204.13807}\ } (\bibinfo {year} {2022})}\BibitemShut {NoStop}%
\bibitem [{\citenamefont {Badue}\ \emph {et~al.}(2021)\citenamefont {Badue},
  \citenamefont {Guidolini}, \citenamefont {Carneiro}, \citenamefont {Azevedo},
  \citenamefont {Cardoso}, \citenamefont {Forechi}, \citenamefont {Jesus},
  \citenamefont {Berriel}, \citenamefont {Paixao}, \citenamefont {Mutz} \emph
  {et~al.}}]{badue2021self}%
  \BibitemOpen
  \bibfield  {author} {\bibinfo {author} {\bibfnamefont {C.}~\bibnamefont
  {Badue}}, \bibinfo {author} {\bibfnamefont {R.}~\bibnamefont {Guidolini}},
  \bibinfo {author} {\bibfnamefont {R.~V.}\ \bibnamefont {Carneiro}}, \bibinfo
  {author} {\bibfnamefont {P.}~\bibnamefont {Azevedo}}, \bibinfo {author}
  {\bibfnamefont {V.~B.}\ \bibnamefont {Cardoso}}, \bibinfo {author}
  {\bibfnamefont {A.}~\bibnamefont {Forechi}}, \bibinfo {author} {\bibfnamefont
  {L.}~\bibnamefont {Jesus}}, \bibinfo {author} {\bibfnamefont
  {R.}~\bibnamefont {Berriel}}, \bibinfo {author} {\bibfnamefont {T.~M.}\
  \bibnamefont {Paixao}}, \bibinfo {author} {\bibfnamefont {F.}~\bibnamefont
  {Mutz}}, \emph {et~al.},\ }\bibfield  {title} {\bibinfo {title} {Self-driving
  cars: A survey},\ }\href@noop {} {\bibfield  {journal} {\bibinfo  {journal}
  {Expert Systems with Applications}\ }\textbf {\bibinfo {volume} {165}},\
  \bibinfo {pages} {113816} (\bibinfo {year} {2021})}\BibitemShut {NoStop}%
\bibitem [{\citenamefont {Huang}\ \emph {et~al.}(2021)\citenamefont {Huang},
  \citenamefont {Broughton}, \citenamefont {Mohseni}, \citenamefont {Babbush},
  \citenamefont {Boixo}, \citenamefont {Neven},\ and\ \citenamefont
  {McClean}}]{huang2021power}%
  \BibitemOpen
  \bibfield  {author} {\bibinfo {author} {\bibfnamefont {H.-Y.}\ \bibnamefont
  {Huang}}, \bibinfo {author} {\bibfnamefont {M.}~\bibnamefont {Broughton}},
  \bibinfo {author} {\bibfnamefont {M.}~\bibnamefont {Mohseni}}, \bibinfo
  {author} {\bibfnamefont {R.}~\bibnamefont {Babbush}}, \bibinfo {author}
  {\bibfnamefont {S.}~\bibnamefont {Boixo}}, \bibinfo {author} {\bibfnamefont
  {H.}~\bibnamefont {Neven}},\ and\ \bibinfo {author} {\bibfnamefont {J.~R.}\
  \bibnamefont {McClean}},\ }\bibfield  {title} {\bibinfo {title} {Power of
  data in quantum machine learning},\ }\href@noop {} {\bibfield  {journal}
  {\bibinfo  {journal} {Nature communications}\ }\textbf {\bibinfo {volume}
  {12}},\ \bibinfo {pages} {1} (\bibinfo {year} {2021})}\BibitemShut {NoStop}%
\bibitem [{\citenamefont {Farhi}\ \emph {et~al.}(2001)\citenamefont {Farhi},
  \citenamefont {Goldstone}, \citenamefont {Gutmann}, \citenamefont {Lapan},
  \citenamefont {Lundgren},\ and\ \citenamefont {Preda}}]{farhi2001quantum}%
  \BibitemOpen
  \bibfield  {author} {\bibinfo {author} {\bibfnamefont {E.}~\bibnamefont
  {Farhi}}, \bibinfo {author} {\bibfnamefont {J.}~\bibnamefont {Goldstone}},
  \bibinfo {author} {\bibfnamefont {S.}~\bibnamefont {Gutmann}}, \bibinfo
  {author} {\bibfnamefont {J.}~\bibnamefont {Lapan}}, \bibinfo {author}
  {\bibfnamefont {A.}~\bibnamefont {Lundgren}},\ and\ \bibinfo {author}
  {\bibfnamefont {D.}~\bibnamefont {Preda}},\ }\bibfield  {title} {\bibinfo
  {title} {A quantum adiabatic evolution algorithm applied to random instances
  of an np-complete problem},\ }\href@noop {} {\bibfield  {journal} {\bibinfo
  {journal} {Science}\ }\textbf {\bibinfo {volume} {292}},\ \bibinfo {pages}
  {472} (\bibinfo {year} {2001})}\BibitemShut {NoStop}%
\bibitem [{\citenamefont {Durr}\ and\ \citenamefont
  {Hoyer}(1996)}]{durr1996quantum}%
  \BibitemOpen
  \bibfield  {author} {\bibinfo {author} {\bibfnamefont {C.}~\bibnamefont
  {Durr}}\ and\ \bibinfo {author} {\bibfnamefont {P.}~\bibnamefont {Hoyer}},\
  }\bibfield  {title} {\bibinfo {title} {A quantum algorithm for finding the
  minimum},\ }\href@noop {} {\bibfield  {journal} {\bibinfo  {journal} {arXiv
  preprint quant-ph/9607014}\ } (\bibinfo {year} {1996})}\BibitemShut {NoStop}%
\bibitem [{\citenamefont {Rebentrost}\ \emph {et~al.}(2014)\citenamefont
  {Rebentrost}, \citenamefont {Mohseni},\ and\ \citenamefont
  {Lloyd}}]{rebentrost2014quantum}%
  \BibitemOpen
  \bibfield  {author} {\bibinfo {author} {\bibfnamefont {P.}~\bibnamefont
  {Rebentrost}}, \bibinfo {author} {\bibfnamefont {M.}~\bibnamefont
  {Mohseni}},\ and\ \bibinfo {author} {\bibfnamefont {S.}~\bibnamefont
  {Lloyd}},\ }\bibfield  {title} {\bibinfo {title} {Quantum support vector
  machine for big data classification},\ }\href@noop {} {\bibfield  {journal}
  {\bibinfo  {journal} {Physical review letters}\ }\textbf {\bibinfo {volume}
  {113}},\ \bibinfo {pages} {130503} (\bibinfo {year} {2014})}\BibitemShut
  {NoStop}%
\bibitem [{\citenamefont {Harrow}\ \emph {et~al.}(2009)\citenamefont {Harrow},
  \citenamefont {Hassidim},\ and\ \citenamefont {Lloyd}}]{harrow2009quantum}%
  \BibitemOpen
  \bibfield  {author} {\bibinfo {author} {\bibfnamefont {A.~W.}\ \bibnamefont
  {Harrow}}, \bibinfo {author} {\bibfnamefont {A.}~\bibnamefont {Hassidim}},\
  and\ \bibinfo {author} {\bibfnamefont {S.}~\bibnamefont {Lloyd}},\ }\bibfield
   {title} {\bibinfo {title} {Quantum algorithm for linear systems of
  equations},\ }\href@noop {} {\bibfield  {journal} {\bibinfo  {journal}
  {Physical review letters}\ }\textbf {\bibinfo {volume} {103}},\ \bibinfo
  {pages} {150502} (\bibinfo {year} {2009})}\BibitemShut {NoStop}%
\bibitem [{\citenamefont {Grover}(1996)}]{grover1996fast}%
  \BibitemOpen
  \bibfield  {author} {\bibinfo {author} {\bibfnamefont {L.~K.}\ \bibnamefont
  {Grover}},\ }\bibfield  {title} {\bibinfo {title} {A fast quantum mechanical
  algorithm for database search},\ }in\ \href@noop {} {\emph {\bibinfo
  {booktitle} {Proceedings of the twenty-eighth annual ACM symposium on Theory
  of computing}}}\ (\bibinfo {year} {1996})\ pp.\ \bibinfo {pages}
  {212--219}\BibitemShut {NoStop}%
\bibitem [{\citenamefont {Neven}\ \emph {et~al.}(2009)\citenamefont {Neven},
  \citenamefont {Denchev}, \citenamefont {Rose},\ and\ \citenamefont
  {Macready}}]{neven2009training}%
  \BibitemOpen
  \bibfield  {author} {\bibinfo {author} {\bibfnamefont {H.}~\bibnamefont
  {Neven}}, \bibinfo {author} {\bibfnamefont {V.~S.}\ \bibnamefont {Denchev}},
  \bibinfo {author} {\bibfnamefont {G.}~\bibnamefont {Rose}},\ and\ \bibinfo
  {author} {\bibfnamefont {W.~G.}\ \bibnamefont {Macready}},\ }\bibfield
  {title} {\bibinfo {title} {Training a large scale classifier with the quantum
  adiabatic algorithm},\ }\href@noop {} {\bibfield  {journal} {\bibinfo
  {journal} {arXiv preprint arXiv:0912.0779}\ } (\bibinfo {year}
  {2009})}\BibitemShut {NoStop}%
\bibitem [{\citenamefont {Leifer}\ and\ \citenamefont
  {Poulin}(2008)}]{leifer2008quantum}%
  \BibitemOpen
  \bibfield  {author} {\bibinfo {author} {\bibfnamefont {M.~S.}\ \bibnamefont
  {Leifer}}\ and\ \bibinfo {author} {\bibfnamefont {D.}~\bibnamefont
  {Poulin}},\ }\bibfield  {title} {\bibinfo {title} {Quantum graphical models
  and belief propagation},\ }\href@noop {} {\bibfield  {journal} {\bibinfo
  {journal} {Annals of Physics}\ }\textbf {\bibinfo {volume} {323}},\ \bibinfo
  {pages} {1899} (\bibinfo {year} {2008})}\BibitemShut {NoStop}%
\bibitem [{\citenamefont {McClean}\ \emph {et~al.}(2021)\citenamefont
  {McClean}, \citenamefont {Harrigan}, \citenamefont {Mohseni}, \citenamefont
  {Rubin}, \citenamefont {Jiang}, \citenamefont {Boixo}, \citenamefont
  {Smelyanskiy}, \citenamefont {Babbush},\ and\ \citenamefont
  {Neven}}]{mcclean2021low}%
  \BibitemOpen
  \bibfield  {author} {\bibinfo {author} {\bibfnamefont {J.~R.}\ \bibnamefont
  {McClean}}, \bibinfo {author} {\bibfnamefont {M.~P.}\ \bibnamefont
  {Harrigan}}, \bibinfo {author} {\bibfnamefont {M.}~\bibnamefont {Mohseni}},
  \bibinfo {author} {\bibfnamefont {N.~C.}\ \bibnamefont {Rubin}}, \bibinfo
  {author} {\bibfnamefont {Z.}~\bibnamefont {Jiang}}, \bibinfo {author}
  {\bibfnamefont {S.}~\bibnamefont {Boixo}}, \bibinfo {author} {\bibfnamefont
  {V.~N.}\ \bibnamefont {Smelyanskiy}}, \bibinfo {author} {\bibfnamefont
  {R.}~\bibnamefont {Babbush}},\ and\ \bibinfo {author} {\bibfnamefont
  {H.}~\bibnamefont {Neven}},\ }\bibfield  {title} {\bibinfo {title} {Low-depth
  mechanisms for quantum optimization},\ }\href@noop {} {\bibfield  {journal}
  {\bibinfo  {journal} {PRX Quantum}\ }\textbf {\bibinfo {volume} {2}},\
  \bibinfo {pages} {030312} (\bibinfo {year} {2021})}\BibitemShut {NoStop}%
\bibitem [{\citenamefont {Aaronson}\ and\ \citenamefont
  {Ambainis}(2009)}]{aaronson2009need}%
  \BibitemOpen
  \bibfield  {author} {\bibinfo {author} {\bibfnamefont {S.}~\bibnamefont
  {Aaronson}}\ and\ \bibinfo {author} {\bibfnamefont {A.}~\bibnamefont
  {Ambainis}},\ }\bibfield  {title} {\bibinfo {title} {The need for structure
  in quantum speedups},\ }\href@noop {} {\bibfield  {journal} {\bibinfo
  {journal} {arXiv preprint arXiv:0911.0996}\ } (\bibinfo {year}
  {2009})}\BibitemShut {NoStop}%
\bibitem [{\citenamefont {Arute}\ \emph {et~al.}(2019)\citenamefont {Arute},
  \citenamefont {Arya}, \citenamefont {Babbush}, \citenamefont {Bacon},
  \citenamefont {Bardin}, \citenamefont {Barends}, \citenamefont {Biswas},
  \citenamefont {Boixo}, \citenamefont {Brandao}, \citenamefont {Buell} \emph
  {et~al.}}]{arute2019quantum}%
  \BibitemOpen
  \bibfield  {author} {\bibinfo {author} {\bibfnamefont {F.}~\bibnamefont
  {Arute}}, \bibinfo {author} {\bibfnamefont {K.}~\bibnamefont {Arya}},
  \bibinfo {author} {\bibfnamefont {R.}~\bibnamefont {Babbush}}, \bibinfo
  {author} {\bibfnamefont {D.}~\bibnamefont {Bacon}}, \bibinfo {author}
  {\bibfnamefont {J.~C.}\ \bibnamefont {Bardin}}, \bibinfo {author}
  {\bibfnamefont {R.}~\bibnamefont {Barends}}, \bibinfo {author} {\bibfnamefont
  {R.}~\bibnamefont {Biswas}}, \bibinfo {author} {\bibfnamefont
  {S.}~\bibnamefont {Boixo}}, \bibinfo {author} {\bibfnamefont {F.~G.}\
  \bibnamefont {Brandao}}, \bibinfo {author} {\bibfnamefont {D.~A.}\
  \bibnamefont {Buell}}, \emph {et~al.},\ }\bibfield  {title} {\bibinfo {title}
  {Quantum supremacy using a programmable superconducting processor},\
  }\href@noop {} {\bibfield  {journal} {\bibinfo  {journal} {Nature}\ }\textbf
  {\bibinfo {volume} {574}},\ \bibinfo {pages} {505} (\bibinfo {year}
  {2019})}\BibitemShut {NoStop}%
\bibitem [{\citenamefont {Boixo}\ \emph {et~al.}(2019)\citenamefont {Boixo},
  \citenamefont {Brandao}, \citenamefont {Buell} \emph
  {et~al.}}]{boixo2019quantum}%
  \BibitemOpen
  \bibfield  {author} {\bibinfo {author} {\bibfnamefont {S.}~\bibnamefont
  {Boixo}}, \bibinfo {author} {\bibfnamefont {F.}~\bibnamefont {Brandao}},
  \bibinfo {author} {\bibfnamefont {D.}~\bibnamefont {Buell}}, \emph {et~al.},\
  }\bibfield  {title} {\bibinfo {title} {Quantum supremacy using a programmable
  superconducting processor},\ }\href@noop {} {\bibfield  {journal} {\bibinfo
  {journal} {Nature}\ }\textbf {\bibinfo {volume} {574}},\ \bibinfo {pages}
  {505} (\bibinfo {year} {2019})}\BibitemShut {NoStop}%
\bibitem [{\citenamefont {McClean}\ \emph {et~al.}(2016)\citenamefont
  {McClean}, \citenamefont {Romero}, \citenamefont {Babbush},\ and\
  \citenamefont {Aspuru-Guzik}}]{mcclean2016theory}%
  \BibitemOpen
  \bibfield  {author} {\bibinfo {author} {\bibfnamefont {J.~R.}\ \bibnamefont
  {McClean}}, \bibinfo {author} {\bibfnamefont {J.}~\bibnamefont {Romero}},
  \bibinfo {author} {\bibfnamefont {R.}~\bibnamefont {Babbush}},\ and\ \bibinfo
  {author} {\bibfnamefont {A.}~\bibnamefont {Aspuru-Guzik}},\ }\bibfield
  {title} {\bibinfo {title} {The theory of variational hybrid quantum-classical
  algorithms},\ }\href@noop {} {\bibfield  {journal} {\bibinfo  {journal} {New
  Journal of Physics}\ }\textbf {\bibinfo {volume} {18}},\ \bibinfo {pages}
  {023023} (\bibinfo {year} {2016})}\BibitemShut {NoStop}%
\bibitem [{\citenamefont {Farhi}\ and\ \citenamefont
  {Neven}(2018)}]{farhi2018classification}%
  \BibitemOpen
  \bibfield  {author} {\bibinfo {author} {\bibfnamefont {E.}~\bibnamefont
  {Farhi}}\ and\ \bibinfo {author} {\bibfnamefont {H.}~\bibnamefont {Neven}},\
  }\bibfield  {title} {\bibinfo {title} {Classification with quantum neural
  networks on near term processors},\ }\href@noop {} {\bibfield  {journal}
  {\bibinfo  {journal} {arXiv preprint arXiv:1802.06002}\ } (\bibinfo {year}
  {2018})}\BibitemShut {NoStop}%
\bibitem [{\citenamefont {Peruzzo}\ \emph {et~al.}(2014)\citenamefont
  {Peruzzo}, \citenamefont {McClean}, \citenamefont {Shadbolt}, \citenamefont
  {Yung}, \citenamefont {Zhou}, \citenamefont {Love}, \citenamefont
  {Aspuru-Guzik},\ and\ \citenamefont {O’brien}}]{peruzzo2014variational}%
  \BibitemOpen
  \bibfield  {author} {\bibinfo {author} {\bibfnamefont {A.}~\bibnamefont
  {Peruzzo}}, \bibinfo {author} {\bibfnamefont {J.}~\bibnamefont {McClean}},
  \bibinfo {author} {\bibfnamefont {P.}~\bibnamefont {Shadbolt}}, \bibinfo
  {author} {\bibfnamefont {M.-H.}\ \bibnamefont {Yung}}, \bibinfo {author}
  {\bibfnamefont {X.-Q.}\ \bibnamefont {Zhou}}, \bibinfo {author}
  {\bibfnamefont {P.~J.}\ \bibnamefont {Love}}, \bibinfo {author}
  {\bibfnamefont {A.}~\bibnamefont {Aspuru-Guzik}},\ and\ \bibinfo {author}
  {\bibfnamefont {J.~L.}\ \bibnamefont {O’brien}},\ }\bibfield  {title}
  {\bibinfo {title} {A variational eigenvalue solver on a photonic quantum
  processor},\ }\href@noop {} {\bibfield  {journal} {\bibinfo  {journal}
  {Nature communications}\ }\textbf {\bibinfo {volume} {5}},\ \bibinfo {pages}
  {1} (\bibinfo {year} {2014})}\BibitemShut {NoStop}%
\bibitem [{\citenamefont {McClean}\ \emph {et~al.}(2018)\citenamefont
  {McClean}, \citenamefont {Boixo}, \citenamefont {Smelyanskiy}, \citenamefont
  {Babbush},\ and\ \citenamefont {Neven}}]{mcclean2018barren}%
  \BibitemOpen
  \bibfield  {author} {\bibinfo {author} {\bibfnamefont {J.~R.}\ \bibnamefont
  {McClean}}, \bibinfo {author} {\bibfnamefont {S.}~\bibnamefont {Boixo}},
  \bibinfo {author} {\bibfnamefont {V.~N.}\ \bibnamefont {Smelyanskiy}},
  \bibinfo {author} {\bibfnamefont {R.}~\bibnamefont {Babbush}},\ and\ \bibinfo
  {author} {\bibfnamefont {H.}~\bibnamefont {Neven}},\ }\bibfield  {title}
  {\bibinfo {title} {Barren plateaus in quantum neural network training
  landscapes},\ }\href@noop {} {\bibfield  {journal} {\bibinfo  {journal}
  {Nature communications}\ }\textbf {\bibinfo {volume} {9}},\ \bibinfo {pages}
  {4812} (\bibinfo {year} {2018})}\BibitemShut {NoStop}%
\bibitem [{\citenamefont {Cerezo}\ \emph {et~al.}(2021)\citenamefont {Cerezo},
  \citenamefont {Sone}, \citenamefont {Volkoff}, \citenamefont {Cincio},\ and\
  \citenamefont {Coles}}]{cerezo2021cost}%
  \BibitemOpen
  \bibfield  {author} {\bibinfo {author} {\bibfnamefont {M.}~\bibnamefont
  {Cerezo}}, \bibinfo {author} {\bibfnamefont {A.}~\bibnamefont {Sone}},
  \bibinfo {author} {\bibfnamefont {T.}~\bibnamefont {Volkoff}}, \bibinfo
  {author} {\bibfnamefont {L.}~\bibnamefont {Cincio}},\ and\ \bibinfo {author}
  {\bibfnamefont {P.~J.}\ \bibnamefont {Coles}},\ }\bibfield  {title} {\bibinfo
  {title} {Cost function dependent barren plateaus in shallow parametrized
  quantum circuits},\ }\href@noop {} {\bibfield  {journal} {\bibinfo  {journal}
  {Nature communications}\ }\textbf {\bibinfo {volume} {12}},\ \bibinfo {pages}
  {1791} (\bibinfo {year} {2021})}\BibitemShut {NoStop}%
\bibitem [{\citenamefont {Anschuetz}\ and\ \citenamefont
  {Kiani}(2022)}]{anschuetz2022quantum}%
  \BibitemOpen
  \bibfield  {author} {\bibinfo {author} {\bibfnamefont {E.~R.}\ \bibnamefont
  {Anschuetz}}\ and\ \bibinfo {author} {\bibfnamefont {B.~T.}\ \bibnamefont
  {Kiani}},\ }\bibfield  {title} {\bibinfo {title} {Quantum variational
  algorithms are swamped with traps},\ }\href@noop {} {\bibfield  {journal}
  {\bibinfo  {journal} {Nature Communications}\ }\textbf {\bibinfo {volume}
  {13}},\ \bibinfo {pages} {7760} (\bibinfo {year} {2022})}\BibitemShut
  {NoStop}%
\bibitem [{\citenamefont {Lloyd}\ \emph {et~al.}(2014)\citenamefont {Lloyd},
  \citenamefont {Mohseni},\ and\ \citenamefont
  {Rebentrost}}]{lloyd2014quantum}%
  \BibitemOpen
  \bibfield  {author} {\bibinfo {author} {\bibfnamefont {S.}~\bibnamefont
  {Lloyd}}, \bibinfo {author} {\bibfnamefont {M.}~\bibnamefont {Mohseni}},\
  and\ \bibinfo {author} {\bibfnamefont {P.}~\bibnamefont {Rebentrost}},\
  }\bibfield  {title} {\bibinfo {title} {Quantum principal component
  analysis},\ }\href@noop {} {\bibfield  {journal} {\bibinfo  {journal} {Nature
  Physics}\ }\textbf {\bibinfo {volume} {10}},\ \bibinfo {pages} {631}
  (\bibinfo {year} {2014})}\BibitemShut {NoStop}%
\bibitem [{\citenamefont {Biamonte}\ \emph {et~al.}(2017)\citenamefont
  {Biamonte}, \citenamefont {Wittek}, \citenamefont {Pancotti}, \citenamefont
  {Rebentrost}, \citenamefont {Wiebe},\ and\ \citenamefont
  {Lloyd}}]{biamonte2017quantum}%
  \BibitemOpen
  \bibfield  {author} {\bibinfo {author} {\bibfnamefont {J.}~\bibnamefont
  {Biamonte}}, \bibinfo {author} {\bibfnamefont {P.}~\bibnamefont {Wittek}},
  \bibinfo {author} {\bibfnamefont {N.}~\bibnamefont {Pancotti}}, \bibinfo
  {author} {\bibfnamefont {P.}~\bibnamefont {Rebentrost}}, \bibinfo {author}
  {\bibfnamefont {N.}~\bibnamefont {Wiebe}},\ and\ \bibinfo {author}
  {\bibfnamefont {S.}~\bibnamefont {Lloyd}},\ }\bibfield  {title} {\bibinfo
  {title} {Quantum machine learning},\ }\href@noop {} {\bibfield  {journal}
  {\bibinfo  {journal} {Nature}\ }\textbf {\bibinfo {volume} {549}},\ \bibinfo
  {pages} {195} (\bibinfo {year} {2017})}\BibitemShut {NoStop}%
\bibitem [{\citenamefont {Wittek}(2014)}]{wittek2014quantum}%
  \BibitemOpen
  \bibfield  {author} {\bibinfo {author} {\bibfnamefont {P.}~\bibnamefont
  {Wittek}},\ }\href@noop {} {\emph {\bibinfo {title} {Quantum machine
  learning: what quantum computing means to data mining}}}\ (\bibinfo
  {publisher} {Academic Press},\ \bibinfo {year} {2014})\BibitemShut {NoStop}%
\bibitem [{\citenamefont {Schuld}(2021)}]{schuld2021supervised}%
  \BibitemOpen
  \bibfield  {author} {\bibinfo {author} {\bibfnamefont {M.}~\bibnamefont
  {Schuld}},\ }\bibfield  {title} {\bibinfo {title} {Supervised quantum machine
  learning models are kernel methods},\ }\href@noop {} {\bibfield  {journal}
  {\bibinfo  {journal} {arXiv preprint arXiv:2101.11020}\ } (\bibinfo {year}
  {2021})}\BibitemShut {NoStop}%
\bibitem [{\citenamefont {A{\"\i}meur}\ \emph {et~al.}(2006)\citenamefont
  {A{\"\i}meur}, \citenamefont {Brassard},\ and\ \citenamefont
  {Gambs}}]{aimeur2006machine}%
  \BibitemOpen
  \bibfield  {author} {\bibinfo {author} {\bibfnamefont {E.}~\bibnamefont
  {A{\"\i}meur}}, \bibinfo {author} {\bibfnamefont {G.}~\bibnamefont
  {Brassard}},\ and\ \bibinfo {author} {\bibfnamefont {S.}~\bibnamefont
  {Gambs}},\ }\bibfield  {title} {\bibinfo {title} {Machine learning in a
  quantum world},\ }in\ \href@noop {} {\emph {\bibinfo {booktitle} {Conference
  of the Canadian Society for Computational Studies of Intelligence}}}\
  (\bibinfo {organization} {Springer},\ \bibinfo {year} {2006})\ pp.\ \bibinfo
  {pages} {431--442}\BibitemShut {NoStop}%
\bibitem [{\citenamefont {Schuld}\ and\ \citenamefont
  {Petruccione}(2018)}]{schuld2018supervised}%
  \BibitemOpen
  \bibfield  {author} {\bibinfo {author} {\bibfnamefont {M.}~\bibnamefont
  {Schuld}}\ and\ \bibinfo {author} {\bibfnamefont {F.}~\bibnamefont
  {Petruccione}},\ }\href@noop {} {\emph {\bibinfo {title} {Supervised learning
  with quantum computers}}},\ Vol.~\bibinfo {volume} {17}\ (\bibinfo
  {publisher} {Springer},\ \bibinfo {year} {2018})\BibitemShut {NoStop}%
\bibitem [{\citenamefont {Schuld}\ and\ \citenamefont
  {Killoran}(2019)}]{schuld2019quantum}%
  \BibitemOpen
  \bibfield  {author} {\bibinfo {author} {\bibfnamefont {M.}~\bibnamefont
  {Schuld}}\ and\ \bibinfo {author} {\bibfnamefont {N.}~\bibnamefont
  {Killoran}},\ }\bibfield  {title} {\bibinfo {title} {Quantum machine learning
  in feature hilbert spaces},\ }\href@noop {} {\bibfield  {journal} {\bibinfo
  {journal} {Physical review letters}\ }\textbf {\bibinfo {volume} {122}},\
  \bibinfo {pages} {040504} (\bibinfo {year} {2019})}\BibitemShut {NoStop}%
\bibitem [{\citenamefont {Havl{\'\i}{\v{c}}ek}\ \emph
  {et~al.}(2019)\citenamefont {Havl{\'\i}{\v{c}}ek}, \citenamefont
  {C{\'o}rcoles}, \citenamefont {Temme}, \citenamefont {Harrow}, \citenamefont
  {Kandala}, \citenamefont {Chow},\ and\ \citenamefont
  {Gambetta}}]{havlivcek2019supervised}%
  \BibitemOpen
  \bibfield  {author} {\bibinfo {author} {\bibfnamefont {V.}~\bibnamefont
  {Havl{\'\i}{\v{c}}ek}}, \bibinfo {author} {\bibfnamefont {A.~D.}\
  \bibnamefont {C{\'o}rcoles}}, \bibinfo {author} {\bibfnamefont
  {K.}~\bibnamefont {Temme}}, \bibinfo {author} {\bibfnamefont {A.~W.}\
  \bibnamefont {Harrow}}, \bibinfo {author} {\bibfnamefont {A.}~\bibnamefont
  {Kandala}}, \bibinfo {author} {\bibfnamefont {J.~M.}\ \bibnamefont {Chow}},\
  and\ \bibinfo {author} {\bibfnamefont {J.~M.}\ \bibnamefont {Gambetta}},\
  }\bibfield  {title} {\bibinfo {title} {Supervised learning with
  quantum-enhanced feature spaces},\ }\href@noop {} {\bibfield  {journal}
  {\bibinfo  {journal} {Nature}\ }\textbf {\bibinfo {volume} {567}},\ \bibinfo
  {pages} {209} (\bibinfo {year} {2019})}\BibitemShut {NoStop}%
\bibitem [{\citenamefont {Schuld}\ \emph {et~al.}(2020)\citenamefont {Schuld},
  \citenamefont {Bocharov}, \citenamefont {Svore},\ and\ \citenamefont
  {Wiebe}}]{schuld2020circuit}%
  \BibitemOpen
  \bibfield  {author} {\bibinfo {author} {\bibfnamefont {M.}~\bibnamefont
  {Schuld}}, \bibinfo {author} {\bibfnamefont {A.}~\bibnamefont {Bocharov}},
  \bibinfo {author} {\bibfnamefont {K.~M.}\ \bibnamefont {Svore}},\ and\
  \bibinfo {author} {\bibfnamefont {N.}~\bibnamefont {Wiebe}},\ }\bibfield
  {title} {\bibinfo {title} {Circuit-centric quantum classifiers},\ }\href@noop
  {} {\bibfield  {journal} {\bibinfo  {journal} {Physical Review A}\ }\textbf
  {\bibinfo {volume} {101}},\ \bibinfo {pages} {032308} (\bibinfo {year}
  {2020})}\BibitemShut {NoStop}%
\bibitem [{\citenamefont {Chatterjee}\ and\ \citenamefont
  {Yu}(2017)}]{chatterjee2017generalized}%
  \BibitemOpen
  \bibfield  {author} {\bibinfo {author} {\bibfnamefont {R.}~\bibnamefont
  {Chatterjee}}\ and\ \bibinfo {author} {\bibfnamefont {T.}~\bibnamefont
  {Yu}},\ }\bibfield  {title} {\bibinfo {title} {Generalized coherent states,
  reproducing kernels, and quantum support vector machines},\ }\href@noop {}
  {\bibfield  {journal} {\bibinfo  {journal} {Quantum Information \&
  Computation}\ }\textbf {\bibinfo {volume} {17}},\ \bibinfo {pages} {1292}
  (\bibinfo {year} {2017})}\BibitemShut {NoStop}%
\bibitem [{\citenamefont {Blank}\ \emph {et~al.}(2020)\citenamefont {Blank},
  \citenamefont {Park}, \citenamefont {Rhee},\ and\ \citenamefont
  {Petruccione}}]{blank2020quantum}%
  \BibitemOpen
  \bibfield  {author} {\bibinfo {author} {\bibfnamefont {C.}~\bibnamefont
  {Blank}}, \bibinfo {author} {\bibfnamefont {D.~K.}\ \bibnamefont {Park}},
  \bibinfo {author} {\bibfnamefont {J.-K.~K.}\ \bibnamefont {Rhee}},\ and\
  \bibinfo {author} {\bibfnamefont {F.}~\bibnamefont {Petruccione}},\
  }\bibfield  {title} {\bibinfo {title} {Quantum classifier with tailored
  quantum kernel},\ }\href@noop {} {\bibfield  {journal} {\bibinfo  {journal}
  {npj Quantum Information}\ }\textbf {\bibinfo {volume} {6}},\ \bibinfo
  {pages} {1} (\bibinfo {year} {2020})}\BibitemShut {NoStop}%
\bibitem [{\citenamefont {Liu}\ and\ \citenamefont
  {Wang}(2018)}]{liu2018differentiable}%
  \BibitemOpen
  \bibfield  {author} {\bibinfo {author} {\bibfnamefont {J.-G.}\ \bibnamefont
  {Liu}}\ and\ \bibinfo {author} {\bibfnamefont {L.}~\bibnamefont {Wang}},\
  }\bibfield  {title} {\bibinfo {title} {Differentiable learning of quantum
  circuit born machines},\ }\href@noop {} {\bibfield  {journal} {\bibinfo
  {journal} {Physical Review A}\ }\textbf {\bibinfo {volume} {98}},\ \bibinfo
  {pages} {062324} (\bibinfo {year} {2018})}\BibitemShut {NoStop}%
\bibitem [{\citenamefont {Liu}\ \emph {et~al.}(2021)\citenamefont {Liu},
  \citenamefont {Arunachalam},\ and\ \citenamefont {Temme}}]{liu2021rigorous}%
  \BibitemOpen
  \bibfield  {author} {\bibinfo {author} {\bibfnamefont {Y.}~\bibnamefont
  {Liu}}, \bibinfo {author} {\bibfnamefont {S.}~\bibnamefont {Arunachalam}},\
  and\ \bibinfo {author} {\bibfnamefont {K.}~\bibnamefont {Temme}},\ }\bibfield
   {title} {\bibinfo {title} {A rigorous and robust quantum speed-up in
  supervised machine learning},\ }\href@noop {} {\bibfield  {journal} {\bibinfo
   {journal} {Nature Physics}\ }\textbf {\bibinfo {volume} {17}},\ \bibinfo
  {pages} {1013} (\bibinfo {year} {2021})}\BibitemShut {NoStop}%
\bibitem [{\citenamefont {K{\"u}bler}\ \emph {et~al.}(2021)\citenamefont
  {K{\"u}bler}, \citenamefont {Buchholz},\ and\ \citenamefont
  {Sch{\"o}lkopf}}]{kubler2021inductive}%
  \BibitemOpen
  \bibfield  {author} {\bibinfo {author} {\bibfnamefont {J.}~\bibnamefont
  {K{\"u}bler}}, \bibinfo {author} {\bibfnamefont {S.}~\bibnamefont
  {Buchholz}},\ and\ \bibinfo {author} {\bibfnamefont {B.}~\bibnamefont
  {Sch{\"o}lkopf}},\ }\bibfield  {title} {\bibinfo {title} {The inductive bias
  of quantum kernels},\ }\href@noop {} {\bibfield  {journal} {\bibinfo
  {journal} {Advances in Neural Information Processing Systems}\ }\textbf
  {\bibinfo {volume} {34}},\ \bibinfo {pages} {12661} (\bibinfo {year}
  {2021})}\BibitemShut {NoStop}%
\bibitem [{\citenamefont {Muthukrishnan}\ \emph {et~al.}(2016)\citenamefont
  {Muthukrishnan}, \citenamefont {Albash},\ and\ \citenamefont
  {Lidar}}]{muthukrishnan2016tunneling}%
  \BibitemOpen
  \bibfield  {author} {\bibinfo {author} {\bibfnamefont {S.}~\bibnamefont
  {Muthukrishnan}}, \bibinfo {author} {\bibfnamefont {T.}~\bibnamefont
  {Albash}},\ and\ \bibinfo {author} {\bibfnamefont {D.~A.}\ \bibnamefont
  {Lidar}},\ }\bibfield  {title} {\bibinfo {title} {Tunneling and speedup in
  quantum optimization for permutation-symmetric problems},\ }\href@noop {}
  {\bibfield  {journal} {\bibinfo  {journal} {Physical Review X}\ }\textbf
  {\bibinfo {volume} {6}},\ \bibinfo {pages} {031010} (\bibinfo {year}
  {2016})}\BibitemShut {NoStop}%
\bibitem [{\citenamefont {Katsuda}\ and\ \citenamefont
  {Nishimori}(2013)}]{katsuda2013nonadiabatic}%
  \BibitemOpen
  \bibfield  {author} {\bibinfo {author} {\bibfnamefont {H.}~\bibnamefont
  {Katsuda}}\ and\ \bibinfo {author} {\bibfnamefont {H.}~\bibnamefont
  {Nishimori}},\ }\bibfield  {title} {\bibinfo {title} {Nonadiabatic quantum
  annealing for one-dimensional trasverse-field ising model},\ }\href@noop {}
  {\bibfield  {journal} {\bibinfo  {journal} {arXiv preprint arXiv:1303.6045}\
  } (\bibinfo {year} {2013})}\BibitemShut {NoStop}%
\bibitem [{\citenamefont {Karanikolas}\ and\ \citenamefont
  {Kawabata}(2020)}]{karanikolas2020pulsed}%
  \BibitemOpen
  \bibfield  {author} {\bibinfo {author} {\bibfnamefont {V.}~\bibnamefont
  {Karanikolas}}\ and\ \bibinfo {author} {\bibfnamefont {S.}~\bibnamefont
  {Kawabata}},\ }\bibfield  {title} {\bibinfo {title} {Pulsed quantum
  annealing},\ }\href@noop {} {\bibfield  {journal} {\bibinfo  {journal}
  {Journal of the Physical Society of Japan}\ }\textbf {\bibinfo {volume}
  {89}},\ \bibinfo {pages} {094003} (\bibinfo {year} {2020})}\BibitemShut
  {NoStop}%
\bibitem [{\citenamefont {Haah}\ \emph {et~al.}(2021)\citenamefont {Haah},
  \citenamefont {Hastings}, \citenamefont {Kothari},\ and\ \citenamefont
  {Low}}]{haah2021quantum}%
  \BibitemOpen
  \bibfield  {author} {\bibinfo {author} {\bibfnamefont {J.}~\bibnamefont
  {Haah}}, \bibinfo {author} {\bibfnamefont {M.~B.}\ \bibnamefont {Hastings}},
  \bibinfo {author} {\bibfnamefont {R.}~\bibnamefont {Kothari}},\ and\ \bibinfo
  {author} {\bibfnamefont {G.~H.}\ \bibnamefont {Low}},\ }\bibfield  {title}
  {\bibinfo {title} {Quantum algorithm for simulating real time evolution of
  lattice hamiltonians},\ }\href@noop {} {\bibfield  {journal} {\bibinfo
  {journal} {SIAM Journal on Computing}\ ,\ \bibinfo {pages} {FOCS18}}
  (\bibinfo {year} {2021})}\BibitemShut {NoStop}%
\bibitem [{\citenamefont {Krizhevsky}\ \emph {et~al.}(2009)\citenamefont
  {Krizhevsky}, \citenamefont {Hinton} \emph
  {et~al.}}]{krizhevsky2009learning}%
  \BibitemOpen
  \bibfield  {author} {\bibinfo {author} {\bibfnamefont {A.}~\bibnamefont
  {Krizhevsky}}, \bibinfo {author} {\bibfnamefont {G.}~\bibnamefont {Hinton}},
  \emph {et~al.},\ }\bibfield  {title} {\bibinfo {title} {Learning multiple
  layers of features from tiny images},\ }\href@noop {} {\bibfield  {journal}
  {\bibinfo  {journal} {https://www.cs.toronto.edu/~kriz/cifar.html}\ }
  (\bibinfo {year} {2009})}\BibitemShut {NoStop}%
\bibitem [{\citenamefont {Coles}\ \emph {et~al.}(2019)\citenamefont {Coles},
  \citenamefont {Cerezo},\ and\ \citenamefont {Cincio}}]{coles2019strong}%
  \BibitemOpen
  \bibfield  {author} {\bibinfo {author} {\bibfnamefont {P.~J.}\ \bibnamefont
  {Coles}}, \bibinfo {author} {\bibfnamefont {M.}~\bibnamefont {Cerezo}},\ and\
  \bibinfo {author} {\bibfnamefont {L.}~\bibnamefont {Cincio}},\ }\bibfield
  {title} {\bibinfo {title} {Strong bound between trace distance and
  hilbert-schmidt distance for low-rank states},\ }\href@noop {} {\bibfield
  {journal} {\bibinfo  {journal} {Physical Review A}\ }\textbf {\bibinfo
  {volume} {100}},\ \bibinfo {pages} {022103} (\bibinfo {year}
  {2019})}\BibitemShut {NoStop}%
\bibitem [{\citenamefont {Al-Mohy}\ and\ \citenamefont
  {Higham}(2010)}]{al2010new}%
  \BibitemOpen
  \bibfield  {author} {\bibinfo {author} {\bibfnamefont {A.~H.}\ \bibnamefont
  {Al-Mohy}}\ and\ \bibinfo {author} {\bibfnamefont {N.~J.}\ \bibnamefont
  {Higham}},\ }\bibfield  {title} {\bibinfo {title} {A new scaling and squaring
  algorithm for the matrix exponential},\ }\href@noop {} {\bibfield  {journal}
  {\bibinfo  {journal} {SIAM Journal on Matrix Analysis and Applications}\
  }\textbf {\bibinfo {volume} {31}},\ \bibinfo {pages} {970} (\bibinfo {year}
  {2010})}\BibitemShut {NoStop}%
\bibitem [{\citenamefont {Arioli}\ \emph {et~al.}(1996)\citenamefont {Arioli},
  \citenamefont {Codenotti},\ and\ \citenamefont {Fassino}}]{arioli1996pade}%
  \BibitemOpen
  \bibfield  {author} {\bibinfo {author} {\bibfnamefont {M.}~\bibnamefont
  {Arioli}}, \bibinfo {author} {\bibfnamefont {B.}~\bibnamefont {Codenotti}},\
  and\ \bibinfo {author} {\bibfnamefont {C.}~\bibnamefont {Fassino}},\
  }\bibfield  {title} {\bibinfo {title} {The pad{\'e} method for computing the
  matrix exponential},\ }\href@noop {} {\bibfield  {journal} {\bibinfo
  {journal} {Linear algebra and its applications}\ }\textbf {\bibinfo {volume}
  {240}},\ \bibinfo {pages} {111} (\bibinfo {year} {1996})}\BibitemShut
  {NoStop}%
\bibitem [{\citenamefont {An}\ \emph {et~al.}(2020)\citenamefont {An},
  \citenamefont {Lee}, \citenamefont {Park}, \citenamefont {Yang},\ and\
  \citenamefont {So}}]{an2020ensemble}%
  \BibitemOpen
  \bibfield  {author} {\bibinfo {author} {\bibfnamefont {S.}~\bibnamefont
  {An}}, \bibinfo {author} {\bibfnamefont {M.}~\bibnamefont {Lee}}, \bibinfo
  {author} {\bibfnamefont {S.}~\bibnamefont {Park}}, \bibinfo {author}
  {\bibfnamefont {H.}~\bibnamefont {Yang}},\ and\ \bibinfo {author}
  {\bibfnamefont {J.}~\bibnamefont {So}},\ }\bibfield  {title} {\bibinfo
  {title} {An ensemble of simple convolutional neural network models for mnist
  digit recognition},\ }\href@noop {} {\bibfield  {journal} {\bibinfo
  {journal} {arXiv preprint arXiv:2008.10400}\ } (\bibinfo {year}
  {2020})}\BibitemShut {NoStop}%
\bibitem [{\citenamefont {Dosovitskiy}\ \emph {et~al.}(2020)\citenamefont
  {Dosovitskiy}, \citenamefont {Beyer}, \citenamefont {Kolesnikov},
  \citenamefont {Weissenborn}, \citenamefont {Zhai}, \citenamefont
  {Unterthiner}, \citenamefont {Dehghani}, \citenamefont {Minderer},
  \citenamefont {Heigold}, \citenamefont {Gelly} \emph
  {et~al.}}]{dosovitskiy2020image}%
  \BibitemOpen
  \bibfield  {author} {\bibinfo {author} {\bibfnamefont {A.}~\bibnamefont
  {Dosovitskiy}}, \bibinfo {author} {\bibfnamefont {L.}~\bibnamefont {Beyer}},
  \bibinfo {author} {\bibfnamefont {A.}~\bibnamefont {Kolesnikov}}, \bibinfo
  {author} {\bibfnamefont {D.}~\bibnamefont {Weissenborn}}, \bibinfo {author}
  {\bibfnamefont {X.}~\bibnamefont {Zhai}}, \bibinfo {author} {\bibfnamefont
  {T.}~\bibnamefont {Unterthiner}}, \bibinfo {author} {\bibfnamefont
  {M.}~\bibnamefont {Dehghani}}, \bibinfo {author} {\bibfnamefont
  {M.}~\bibnamefont {Minderer}}, \bibinfo {author} {\bibfnamefont
  {G.}~\bibnamefont {Heigold}}, \bibinfo {author} {\bibfnamefont
  {S.}~\bibnamefont {Gelly}}, \emph {et~al.},\ }\bibfield  {title} {\bibinfo
  {title} {An image is worth 16x16 words: Transformers for image recognition at
  scale},\ }\href@noop {} {\bibfield  {journal} {\bibinfo  {journal} {arXiv
  preprint arXiv:2010.11929}\ } (\bibinfo {year} {2020})}\BibitemShut {NoStop}%
\bibitem [{\citenamefont {LeCun}(1998)}]{lecun1998mnist}%
  \BibitemOpen
  \bibfield  {author} {\bibinfo {author} {\bibfnamefont {Y.}~\bibnamefont
  {LeCun}},\ }\bibfield  {title} {\bibinfo {title} {The mnist database of
  handwritten digits},\ }\href@noop {} {\bibfield  {journal} {\bibinfo
  {journal} {http://yann. lecun. com/exdb/mnist/}\ } (\bibinfo {year}
  {1998})}\BibitemShut {NoStop}%
\bibitem [{\citenamefont {Smith}\ and\ \citenamefont
  {Gales}(2001)}]{smith2001speech}%
  \BibitemOpen
  \bibfield  {author} {\bibinfo {author} {\bibfnamefont {N.}~\bibnamefont
  {Smith}}\ and\ \bibinfo {author} {\bibfnamefont {M.}~\bibnamefont {Gales}},\
  }\bibfield  {title} {\bibinfo {title} {Speech recognition using svms},\
  }\href@noop {} {\bibfield  {journal} {\bibinfo  {journal} {Advances in neural
  information processing systems}\ }\textbf {\bibinfo {volume} {14}} (\bibinfo
  {year} {2001})}\BibitemShut {NoStop}%
\bibitem [{\citenamefont {Anguita}\ \emph {et~al.}(2013)\citenamefont
  {Anguita}, \citenamefont {Ghio}, \citenamefont {Oneto}, \citenamefont
  {Parra~Perez},\ and\ \citenamefont {Reyes~Ortiz}}]{anguita2013public}%
  \BibitemOpen
  \bibfield  {author} {\bibinfo {author} {\bibfnamefont {D.}~\bibnamefont
  {Anguita}}, \bibinfo {author} {\bibfnamefont {A.}~\bibnamefont {Ghio}},
  \bibinfo {author} {\bibfnamefont {L.}~\bibnamefont {Oneto}}, \bibinfo
  {author} {\bibfnamefont {X.}~\bibnamefont {Parra~Perez}},\ and\ \bibinfo
  {author} {\bibfnamefont {J.~L.}\ \bibnamefont {Reyes~Ortiz}},\ }\bibfield
  {title} {\bibinfo {title} {A public domain dataset for human activity
  recognition using smartphones},\ }in\ \href@noop {} {\emph {\bibinfo
  {booktitle} {Proceedings of the 21th international European symposium on
  artificial neural networks, computational intelligence and machine
  learning}}}\ (\bibinfo {year} {2013})\ pp.\ \bibinfo {pages}
  {437--442}\BibitemShut {NoStop}%
\bibitem [{\citenamefont {Torlai}\ \emph {et~al.}(2018)\citenamefont {Torlai},
  \citenamefont {Mazzola}, \citenamefont {Carrasquilla}, \citenamefont
  {Troyer}, \citenamefont {Melko},\ and\ \citenamefont
  {Carleo}}]{torlai2018neural}%
  \BibitemOpen
  \bibfield  {author} {\bibinfo {author} {\bibfnamefont {G.}~\bibnamefont
  {Torlai}}, \bibinfo {author} {\bibfnamefont {G.}~\bibnamefont {Mazzola}},
  \bibinfo {author} {\bibfnamefont {J.}~\bibnamefont {Carrasquilla}}, \bibinfo
  {author} {\bibfnamefont {M.}~\bibnamefont {Troyer}}, \bibinfo {author}
  {\bibfnamefont {R.}~\bibnamefont {Melko}},\ and\ \bibinfo {author}
  {\bibfnamefont {G.}~\bibnamefont {Carleo}},\ }\bibfield  {title} {\bibinfo
  {title} {Neural-network quantum state tomography},\ }\href@noop {} {\bibfield
   {journal} {\bibinfo  {journal} {Nature Physics}\ }\textbf {\bibinfo {volume}
  {14}},\ \bibinfo {pages} {447} (\bibinfo {year} {2018})}\BibitemShut
  {NoStop}%
\bibitem [{\citenamefont {Yang}\ \emph {et~al.}(2022)\citenamefont {Yang},
  \citenamefont {Bosch}, \citenamefont {Kiani}, \citenamefont {Lloyd},\ and\
  \citenamefont {Lupascu}}]{yang2022analog}%
  \BibitemOpen
  \bibfield  {author} {\bibinfo {author} {\bibfnamefont {R.}~\bibnamefont
  {Yang}}, \bibinfo {author} {\bibfnamefont {S.}~\bibnamefont {Bosch}},
  \bibinfo {author} {\bibfnamefont {B.}~\bibnamefont {Kiani}}, \bibinfo
  {author} {\bibfnamefont {S.}~\bibnamefont {Lloyd}},\ and\ \bibinfo {author}
  {\bibfnamefont {A.}~\bibnamefont {Lupascu}},\ }\bibfield  {title} {\bibinfo
  {title} {An analog quantum variational embedding classifier},\ }\href@noop {}
  {\bibfield  {journal} {\bibinfo  {journal} {arXiv preprint arXiv:2211.02748}\
  } (\bibinfo {year} {2022})}\BibitemShut {NoStop}%
\bibitem [{\citenamefont {Suzuki}(1991)}]{suzuki1991general}%
  \BibitemOpen
  \bibfield  {author} {\bibinfo {author} {\bibfnamefont {M.}~\bibnamefont
  {Suzuki}},\ }\bibfield  {title} {\bibinfo {title} {General theory of fractal
  path integrals with applications to many-body theories and statistical
  physics},\ }\href@noop {} {\bibfield  {journal} {\bibinfo  {journal} {Journal
  of Mathematical Physics}\ }\textbf {\bibinfo {volume} {32}},\ \bibinfo
  {pages} {400} (\bibinfo {year} {1991})}\BibitemShut {NoStop}%
\bibitem [{\citenamefont {Microsoft}(2022)}]{microsoft2022azure}%
  \BibitemOpen
  \bibfield  {author} {\bibinfo {author} {\bibnamefont {Microsoft}},\
  }\bibfield  {title} {\bibinfo {title} {What is azure quantum?},\ }\href
  {https://docs.microsoft.com/en-us/azure/quantum/} {\bibfield  {journal}
  {\bibinfo  {journal} {Microsoft Azure}\ ,\ \bibinfo {pages} {326}} (\bibinfo
  {year} {2022})}\BibitemShut {NoStop}%
\bibitem [{\citenamefont {Rolnick}\ \emph {et~al.}(2017)\citenamefont
  {Rolnick}, \citenamefont {Veit}, \citenamefont {Belongie},\ and\
  \citenamefont {Shavit}}]{rolnick2017deep}%
  \BibitemOpen
  \bibfield  {author} {\bibinfo {author} {\bibfnamefont {D.}~\bibnamefont
  {Rolnick}}, \bibinfo {author} {\bibfnamefont {A.}~\bibnamefont {Veit}},
  \bibinfo {author} {\bibfnamefont {S.}~\bibnamefont {Belongie}},\ and\
  \bibinfo {author} {\bibfnamefont {N.}~\bibnamefont {Shavit}},\ }\bibfield
  {title} {\bibinfo {title} {Deep learning is robust to massive label noise},\
  }\href@noop {} {\bibfield  {journal} {\bibinfo  {journal} {arXiv preprint
  arXiv:1705.10694}\ } (\bibinfo {year} {2017})}\BibitemShut {NoStop}%
\end{thebibliography}%

\onecolumngrid

\appendix
\section{Implementation in Python}
We implemented a simulation of this setup in \textit{Python}, using \textit{PyTorch}. The source code of the project can be found here \url{https://github.com/BoschSamuel/NNforProgrammingQuantumAnnealers}.

\textit{PyTorch} tensors and \textit{NumPy} arrays are similar in that they both represent multi-dimensional arrays of numerical data. They can be used for similar operations such as indexing, slicing, and mathematical operations. However, there are several key differences between the two. \textit{PyTorch} tensors can be used on a GPU for faster numerical computations, while \textit{NumPy} arrays can only be used on a CPU. Most computations on NNs are highly parallelizable, so that GPUs can help significantly speed up most computations. \textit{PyTorch} also has built-in support for automatic differentiation, which is useful for training machine learning models. In our simulations, we used a hybrid approach in order to optimize for speed while keeping memory usage within our hardware's limits. Since recently, PyTorch tensors also support complex datatypes, essential for performing quantum mechanics calculations. Additionally, \textit{PyTorch} includes a number of additional features, such as support for dynamic computation graphs and a library of pre-defined neural network layers, making it a popular choice for machine learning research and development.

The main challenge in our simulations was memory limitations. We used a computing server with 64 CPUs (2x AMD Ryzen Threadripper PRO 3975WX), 4 GPUs (NVIDIA RTX A6000), and 128 GB of RAM. There exist several different ways of implementing the simulation. Some methods are very fast but require a lot of RAM. Other methods are much slower but require less RAM. The two bottlenecks of our simulations were matrix exponentiations (calculating $e^{\frac{-i \Delta t}{\hslash} H(t_k)}$) and calculating the gradient, which requires storing a new copy of every tensor, every time some of its elements are changed.\\

The problem with matrix exponentiation is that it requires up to $\mathcal{O}(N^3)$ operations, where $N=2^n$. On top of this, the matrix exponentiation has to be performed for every single discretized step, and for every single data point in our dataset. Effectively, this would mean that the total number of operations required for one forward pass in the neural network would be proportional to $\mathcal{O}\Big((8^n) \times (\text{steps}) \times (\text{dataset size})\Big)$. In order to improve this, we make use of the fact that Pauli matrices can be exponentiated:

\begin{equation}\label{eq:pauli_exp}
\begin{split}
    e^{i a\left(\hat{n} \cdot \vec{\sigma}\right)}
    &= \sum_{k=0}^\infty{\frac{i^k \left[a \left(\hat{n} \cdot \vec{\sigma}\right)\right]^k}{k!}} \\
    &= \sum_{p=0}^\infty{\frac{(-1)^p (a\hat{n}\cdot \vec{\sigma})^{2p}}{(2p)!}} + i\sum_{q=0}^\infty{\frac{(-1)^q (a\hat{n}\cdot \vec{\sigma})^{2q + 1}}{(2q + 1)!}} \\
    &= I\sum_{p=0}^\infty{\frac{(-1)^p a^{2p}}{(2p)!}} + i (\hat{n}\cdot \vec{\sigma}) \sum_{q=0}^\infty{\frac{(-1)^q a^{2q+1}}{(2q + 1)!}}\\
    & = I\cos{(a)} + i (\hat{n} \cdot \vec{\sigma}) \sin{(a)}
\end{split}
\end{equation}
where $\vec{\sigma} = (I,\sigma_x,\sigma_y,\sigma_z)^T$.\\

The second method used for speeding up calculations is the Suzuki-Trotter approximation.

\begin{figure*}[htp]
    \centering
    \includegraphics[width=0.6\textwidth]{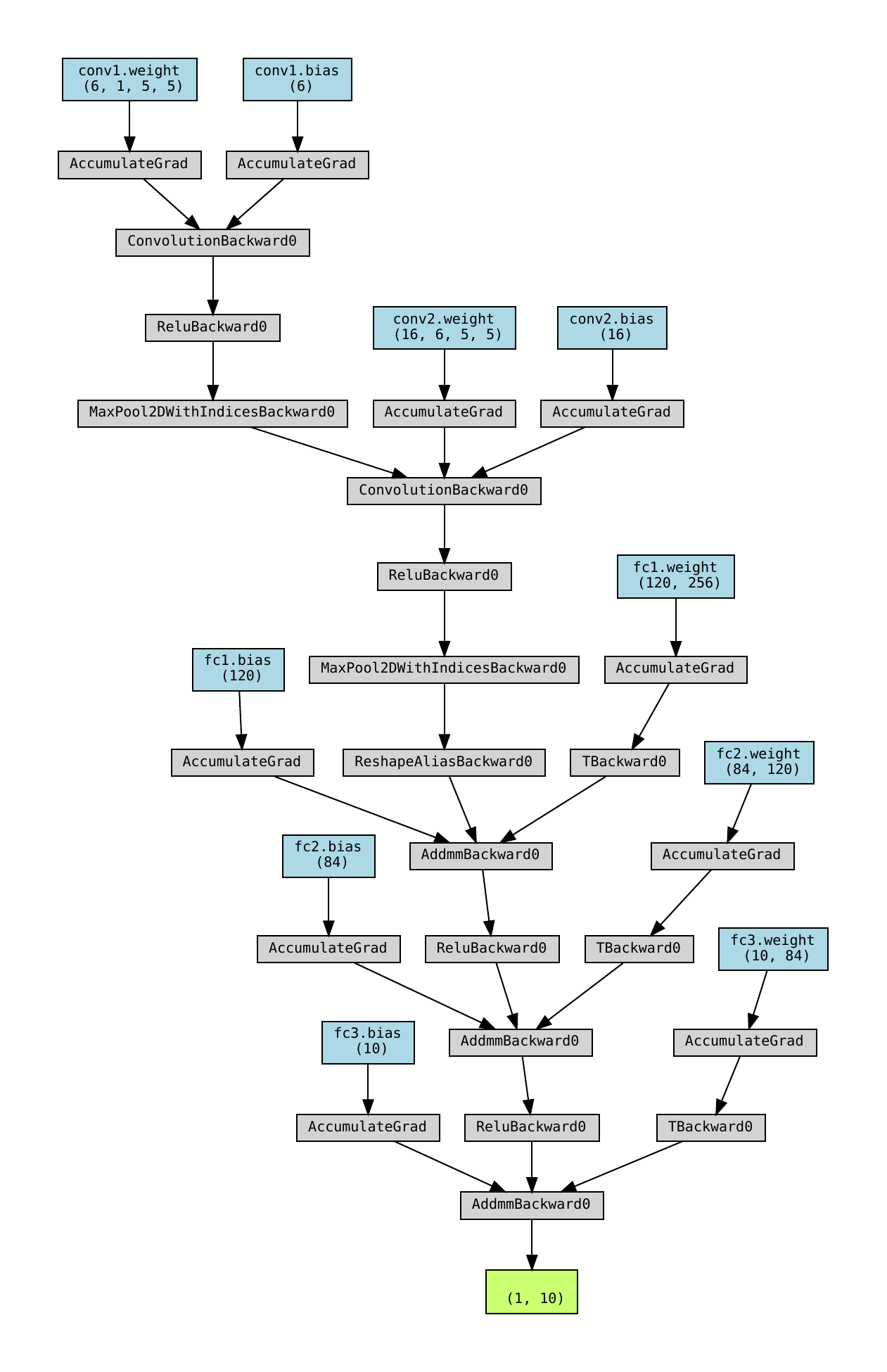}
    \caption{CNN: This is a small classical convolutional neural network consisting of two 2D convolutions. The first convolution has three input channels, six output channels, and a kernel size of five, so it has $(3 * 5 * 5 + 1) * 6 = 456$ parameters. The second convolution has six input channels, $16$ output channels, and a kernel size of five, so it has $(6 * 5 * 5 + 1) * 16 = 2416$ parameters. An input layer with one neuron for each input feature, two hidden layers with $120$ and $84$ neurons, respectively, and an output layer with one neuron for each class. The output is, therefore, $K$ real numbers between $0$ and $1$. For MNIST, $K=10$.}
    \label{fig:CNN}
\end{figure*}

\section{Suzuki-Trotter approximation}\label{sec:trotter}
The Suzuki-Trotter approximation is a technique that can approximate the exponential of a sum of matrices using a product of the exponentials of each matrix. This method can be applied to the exponentiation of Pauli matrices. The Suzuki-Trotter approximation allows for the efficient computation of the exponential of a linear combination of Pauli matrices, which is a crucial operation in many quantum algorithms, including operations performed in this paper. The Suzuki-Trotter method is based on the Trotter-Suzuki formula, which can approximate the exponential of a sum of matrices using a product of exponentials of each individual matrix. The approximation can be controlled by the number of Trotter steps, and the error can be made arbitrarily small by increasing the number of steps.

Concretely, we express the Hamiltonian as a sum of easy-to-simulate Hamiltonians (in our case, linear combinations of Pauli matrices) and then approximate the total evolution as a sequence of these simpler evolutions. Let the Hamiltonian be $H = \sum_{j=1}^m H_j$. For a small $\delta t$, we have:

\begin{equation}
    e^{-i \sum_{j=1}^m H_j \delta t} = \prod_{j=1}^m e^{-i H_j \delta t} + \mathcal{O}(m^2 \delta t^2)
\end{equation}
Here we do \textbf{not} assume that $H_i$ and $H_j$ commute. Otherwise, the above would be an exact expression for any $\delta t$. 

The trick behind the Trotter-Suzuki formula is that this approximation can even be used when, instead of a small $\delta t$, we have a large $t$. This can be approximated by breaking $t$ into $r$ smaller "chunks". The number of "chunks" is called the \textit{Trotter number}, $TN$. More precisely, the Trotter number refers to the number of Trotter steps used in the approximation. The Trotter-Suzuki formula expresses the exponential of a sum of matrices as a product of exponentials of each individual matrix, with an error that decreases as the number of Trotter steps increases. The Trotter number controls the precision of the approximation. As the number of Trotter steps increases, the error in the approximation decreases, and the approximation becomes more accurate. The number of Trotter steps can be chosen based on the desired level of precision and the computational resources available.

\begin{equation}\label{eq:trotter_1st_order}
    e^{-i \sum_{j=1}^m H_j t} = \Bigg( \prod_{j=1}^m e^{-i H_j t/r} \Bigg)^r + \mathcal{O}(m^2 t^2/r)
\end{equation}

Here, in order to keep the error at most $\epsilon$, we have to scale $r$ as $m^2t^2/\epsilon$. In order to achieve even higher accuracy, without having to scale $r$ as much as in equation \ref{eq:trotter_1st_order}, we mostly used the second-order Trotter-Suzuki formula for our simulations:

\begin{equation}\label{eq:trotter_2nd_order}
    e^{-i \sum_{j=1}^m H_j t} = \Bigg( \prod_{j=1}^m e^{-i H_j t/2r} \prod_{j=m}^1 e^{-i H_j t/2r} \Bigg)^r + \mathcal{O}(m^3 t^3/r^2) = U_{2k}(t) + \mathcal{O}(m^3 t^3/r^2)
\end{equation}

Here, in order to keep the error at most $\epsilon$, we have to scale $r$ as $m^{3/2}t^{3/2}/\sqrt{\epsilon}$. It is also possible to get even higher-order formulas by using the recursion formula from (\cite{suzuki1991general}):

\begin{equation}\label{eq:suziki_higher_order}
    U_{2k}(t) = \big[U_{2k-2}(s_k t)  \big]^2 U_{2k - 2} ([1-4s_k] t) [U_{2k-2}(s_k t)]^2 = e^{-i \sum_{j=1}^m H_j t} + \mathcal{O}((mt)^{2k+1}/r^{2k})
\end{equation}

where $s_k = (4 - 4^{1/(2k-1)})^{-1}$

Arbitrarily high-order formulas can be similarly constructed. However, the costs incurred from using more complex integrators often outweigh the benefits beyond the second or fourth order in most practical problems \cite{microsoft2022azure}. Instead, increasing the \textit{Trotter number}, $TN$, should suffice. 

\section{Simulation accuracies}

The \textit{torch.linalg.matrix\_exp} function is highly optimized for parallel computation using GPUs, but its complexity scales roughly as $\mathcal{O}(2^n)^3$. The function uses a Scaling and Squaring algorithm for matrix exponentiation which is based on the Taylor series expansion of the matrix exponential. It starts by scaling the matrix by a power of two, then repeatedly squaring the matrix and summing up the intermediate results. The algorithm's computational complexity is determined by the number of matrix-matrix multiplications, which are the most computationally expensive operations in the algorithm. The number of multiplications is determined by the number of non-zero entries in the matrix to be exponentiated.

\begin{figure}[htp]
    \centering
   \includegraphics[width=0.53\textwidth]{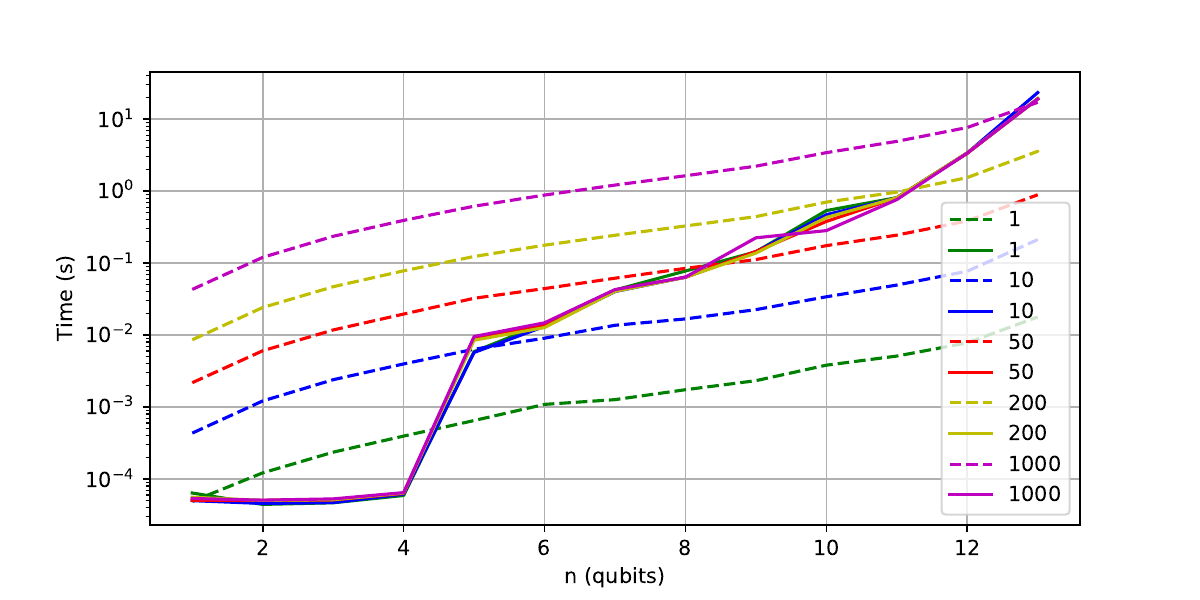}
   \caption{
   Total amount of time for performing the $\ket{\psi({t_{k+1}})} = e^{\frac{-i \Delta t}{\hslash} H(t_k)} \ket{\psi(t_k)}$ update step from equation (\ref{eq:matrix_exponetiation}). This involves single matrix exponentiation of the Hamiltonian $H$ from equation (\ref{eq:Hamiltonian_linear_combination}), and applying (multiplying) it to the previous quantum state. The dotted lines represent times using the Suzuki-Trotter approximation, while the solid lines represent standard matrix exponentiation using \textit{torch.linalg.matrix\_exp} \textbf{and} matrix-vector multiplication using \textit{torch.matmul}. It is important to note that not only the size of the matrix representing this Hamiltonian grows as $2^n$ by $2^n$ or $\mathcal{O}(2^{2n})$, but also the number of terms $H_i$ increases as $\frac{n}{2}(n+3)$ or  $\mathcal{O}(n^2)$. When using the SSA method through \textit{torch.linalg.matrix\_exp}, the number of $H_i$ terms is irrelevant, given that the method works for any arbitrary matrix. Therefore, the function \textit{torch.linalg.matrix\_exp} has a total complexity of $\mathcal{O}(2^{3n})$. On the other hand, using the Suzuki-Trotter approximation allows us to get a complexity scaling of just $\mathcal{O}(n^2 2^{n})$. The reason why this is less than the total number of matrix elements $n^{2n}$ is simply that our implementation of the Suzuki-Trotter approximation mostly relies on symbolic manipulations, as opposed to purely numerical methods. Therefore, the only tensor explicitly stored in the memory is the initial and final state vector of size $2^n$. }
   \label{fig:exp_speed}
\end{figure}

The Suzuki-Trotter approximation involves many individual steps, as well as the implementation of equation (\ref{eq:pauli_exp}) using symbolic operations for gradient conservation purposes. As we can see in figure (\ref{fig:exp_speed}), for $TN=50$, the benefit of using the Suzuki-Trotter approximation only outweighs the standard matrix exponentiation method for $n\geq 9$. Nevertheless, simulations in the range $9 \leq n \leq 15$ are already very slow with either method. 

\begin{figure}[htp]
    \centering
    \includegraphics[width=0.5\textwidth]{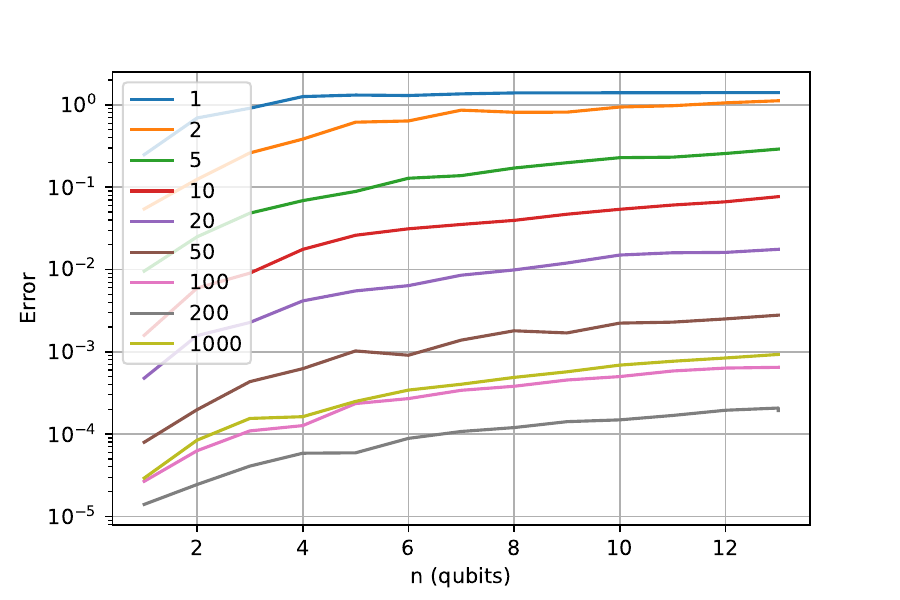}
    \caption{
    The error of the simulations for the entire time-evolution, with $steps=10$. The error is defined as $1-|\braket{\psi_{\text{Trotter approx.}}|\psi_{\text{exact}}}|$. The colors represent different \textit{Trotter numbers} ($TN$). As expected, for a total duration $T=1$ of the simulation, we need an approximation such that $\delta t=\frac{T}{TN} << 1$. The larger $n$, the larger $TN$ has to be, as otherwise, our simulations aren't a good approximation of reality. Interestingly, as our setup includes a NN, the classification accuracy doesn't significantly decrease when the accuracy of our simulation is low. This is because even small NNs are very resilient to noise. \cite{rolnick2017deep} found the same to be true for deep NNs. However, if we went to create a simulation that is reasonably accurate for our classification tasks, we need to pick a $TN$ that is large enough, so the error becomes significantly smaller than the average Hilbert-Schmidt distance between the different classes before the training. Therefore, using $TN\leq50$ should suffice for the purpose of our simulations.
    }
    \label{fig:Error_trotter}
\end{figure}

In order to do so, we ran the simulations of equation (\ref{eq:matrix_exponetiation_enitre_time}) using two different approaches. Firstly, we used the Python function \textit{torch.linalg.matrix\_exp}, based on the Scaling and Squaring approximation (SSA) algorithm, for performing the matrix exponentiation. We named the output state $\ket{\psi_{\text{SSA}}}$. For practical purposes, we can assume that $\ket{\psi_{\text{SSA}}} \approx \ket{\psi_{\text{Exact.}}}$. The second method we used was using the Suzuki-Trotter approximation from section (\ref{sec:trotter}), using different values for Trotter number $TN$ (shown in different colors) and $n$ (the number of qubits). The resulting errors, defined as $\text{Error} = 1-\braket{\psi_{\text{Trotter approx.}}|\psi_{\text{exact}}}$, are shown in figure (\ref{fig:Error_trotter}).

\end{document}